\documentclass[twocolumn,nolinenumbers]{aastex631}

\usepackage{xcolor,tikz}
\usepackage{gensymb}
\usepackage{multirow}

\shorttitle{Fundamental Tests of WD Cooling Physics}
\shortauthors{Barrientos et al.}

\begin{document}

\title{Fundamental Tests of White Dwarf Cooling Physics with Wide Binaries}

\thanks{Author e-mail: \href{mbarrientos@ou.edu}{mbarrientos@ou.edu}}

\author[0000-0002-6153-9304]{Manuel Barrientos} 
\affiliation{Homer L. Dodge Department of Physics and Astronomy, University of Oklahoma, 440 W. Brooks St., Norman OK, 73019, USA}

\author[0000-0001-6098-2235]{Mukremin Kilic} 
\affiliation{Homer L. Dodge Department of Physics and Astronomy, University of Oklahoma, 440 W. Brooks St., Norman OK, 73019, USA}

\author[0000-0003-2368-345X]{Pierre Bergeron} 
\affiliation{Département de Physique, Université de Montréal, C.P. 6128, Succ. Centre-Ville, Montréal, QC H3C 3J7, Canada}

\author[0000-0002-9632-1436]{Simon Blouin}
\affiliation{Department of Physics and Astronomy, University of Victoria, Victoria BC V8W 2Y2, Canada}

\author[0000-0002-4462-2341]{Warren R.\ Brown}
\affiliation{Center for Astrophysics, Harvard \& Smithsonian, 60 Garden Street, Cambridge, MA 02138, USA}

\author[0000-0001-5261-3923]{Jeff J. Andrews}
\affiliation{Department of Physics, University of Florida, 2001 Museum Rd, Gainesville, FL 32611, USA}

\begin{abstract}
We present follow-up spectroscopy and a detailed model atmosphere analysis of 29 wide double white dwarfs, including eight systems with a crystallized C/O core member. We use state-of-the-art evolutionary models to constrain the physical parameters of each star, including the total age. Assuming that the members of wide binaries are coeval, any age difference between the binary members can be used to test the cooling physics for white dwarf stars, including potential delays due to crystallization and $^{22}$Ne distillation. We use our control sample of 14 wide binaries with non-crystallized members to show that this method works well; the control sample shows an age difference of only $\Delta$Age = $-0.03 \pm$ 0.15 Gyr between its members. For the eight crystallized C/O core systems we find a cooling anomaly of $\Delta$Age= 1.13$^{+1.20}_{-1.07}$ Gyr. Even though our results are consistent with a small additional cooling delay ($\sim1$ Gyr) from $^{22}$Ne distillation and other neutron-rich impurities, the large uncertainties make this result not statistically significant. Nevertheless, we rule out cooling delays longer than 3.6 Gyr at the 99.7\% ($3\sigma$) confidence level for 0.6-0.9 $M_{\odot}$ white dwarfs. Further progress requires larger samples of wide binaries with crystallized massive white dwarf members. We provide a list of subgiant + white dwarf binaries that could be used for this purpose in the future.
\end{abstract}

 
\section{Introduction} 
\label{sec:1}
Stars with initial masses below 8-10 M$_{\odot}$ expel a significant fraction of their masses to the interstellar medium, leaving behind a white dwarf (WD) as the remnant. These objects typically consist of a C/O core supported by electron degeneracy pressure that constitutes 99\% of the mass and surface layers of He and H that account for the remaining 1\% \citep[e.g.,][]{saumon2022}. As a WD cools \citep{mestel1952}, electrostatic interactions become more important than the thermal motion of the ions, leading to the formation of a lattice structure in the center of the star, a process that we know as core crystallization \citep{kirzhnits60,abrikosov60,salpeter61}. This first-order transition from liquid to solid phase releases a considerable amount of latent heat. The extra energy slows down the cooling rate of the WD, resulting in a crystallization sequence in the Hertzsprung-Russell (H-R) diagram \citep{vanhorn1968}. \citet{winget09} found indirect evidence for crystallization in the WD cooling sequence of the globular cluster NGC 6397, while \citet{garciaberro2010} invoked additional cooling delays from $^{22}$Ne sedimentation and C/O phase separation upon crystallization to explain the cooling sequence of the metal-rich open cluster NGC 6791,  but direct evidence had to wait another decade. 

Thanks to the precise astrometric data from {\emph Gaia} \citep{gaiadr2}, it is now possible to create volume-limited WD samples that provide unbiased estimates of the luminosity and mass functions. Using the {\emph Gaia} H-R diagram for WDs, \citet{tremblay2019} found that the crystallization sequence is a mass-dependent pile-up in the {\emph Gaia} H-R diagram due to the release of latent heat. In addition, they also found strong evidence for additional cooling delays due to element sedimentation and proposed $^{16}$O sedimentation as a potential source of these extra delays. 

Several authors have studied additional mechanisms for extra energy release related to crystallization and its associated effects. \citet{cheng2019} investigated the properties of WDs in the so-called Q-branch, located in the high-mass end of the WD crystallization sequence, and found that about 6\% of the high-mass WDs must experience an $\geq$8 Gyr extra cooling delay on the Q branch. They suggested $^{22}$Ne settling in the liquid cores of C/O WDs as a source of this extra cooling delay. However, this scenario requires relatively large amounts of $^{22}$Ne (6\% by mass) to reproduce the observed delay \citep[see Figure 4 in][]{camisassa2021}. 

Instead, \citet{blouin2021a} proposed that the phase separation of $^{22}$Ne in a crystallizing C/O WD can lead to a distillation process. Distillation efficiently transports $^{22}$Ne towards the center, releasing a considerable amount of gravitational energy. Under a standard WD composition with X($^{22}$Ne)= 0.014 and X(O)= 0.60, the phase separation process releases energy after the WD is $\sim60$\% crystallized, creating a cooling delay of $\sim$1 Gyr for typical $M=0.6~M_{\odot}$ WDs. \citet{blouin2021a} also argued that this mechanism can largely resolve the ultra-massive cooling anomaly if the delayed population consists of WDs with moderately above-average $^{22}$Ne abundances, X($^{22}$Ne)= 0.03. Therefore, $^{22}$Ne distillation is a promising solution to the cooling anomaly seen in the {\emph Gaia} H-R diagram \citep{bedard24}.

Although these new physics and observations of the luminosity function are strong tools to test the WD cooling models, the simulations in these works are sensitive to model assumptions such as the star formation rate in our Galaxy \citep[see, e.g.,][]{tremblay2014,tremblay2019,fantin2019,mor2019,alzate2021,fleury2022}. Another unknown is the core composition, specifically the C/O ratio, which is still poorly constrained \citep{giammi2018}. The C/O ratio has a significant impact on the amount of energy released during the phase separation of $^{22}$Ne \citep{blouin2021a}.

To avoid population issues, a more direct approach is to study and test the evolutionary models through ages of individual WDs that have gone through crystallization \citep{venner2023}. The ages of isolated WDs cannot be constrained directly. However, if a WD is located in a star cluster, binary, or multiple-star system, its age can be measured through the other stars in the system. These groups of stars usually contain at least one non-degenerate component from which we can infer the total age of the WDs \citep[e.g.,][]{catalan2008,cummings2018,barrientos2021}. \citet{venner2023} identified a Sirius-like quadruple system HD 190412 composed of a crystallizing WD and a non-degenerate triple. The WD (HD 190412 C) is $\approx$65\% crystallized considering a homogeneous C/O core with an O mass fraction of X(O)=0.60, and its age is 4.2$\pm$0.2 Gyr, assuming a metallicity of [Fe/H]= $-$0.25. Comparing this value with the system age of 7.3$^{+1.9}_{-1.8} $ Gyr, they found an age anomaly of +3.1$\pm$1.9 Gyr. The precision of this value is too low to make any statistically significant conclusions about cooling delays from crystallization and its associated effects. However, their findings are consistent with the amplitude of cooling delays expected from $^{22}$Ne distillation.

Wide double white dwarf binaries (DWDs) are nature's smallest star clusters. They have traditionally been used to constrain the initial-final mass relation (IFMR) that connects the progenitor main-sequence mass with the final WD remnant mass \citep{andrews2015,holland2024}. In these systems, the cooling ages, crystallization fractions, and masses of both WDs can be derived using the appropriate spectroscopic and photometric information and the evolutionary models. 

In this paper, we aim to constrain the cooling delays in WDs due to crystallization and its associated effects by using DWDs where one of the members is crystallized. Using an IFMR obtained from young and hot WDs in open clusters \citep[e.g.,][]{cummings2018,williams2018}, we calculate the progenitor mass and lifetime of both components. Since the open cluster WDs are relatively hot and young, they do not suffer from the cooling delays associated with crystallization, and therefore, the resulting IFMR does not suffer from the current problems in our understanding of cooling physics. Hence, the total age of the non-crystallized component can be measured reliably as long as the WD mass is above $0.63~M_{\odot}$ \citep[][]{heintz2022}, providing a benchmark for the whole system. By analyzing the differences in ages between the crystallized and non-crystallized components ($\Delta$Age), we can provide empirical constraints on the cooling delays associated with crystallization and distillation. 

We describe our sample selection in Section~\ref{sec:2}. In Section~\ref{sec:3}, we present the determination of the atmospheric parameters along with the Bayesian ages for all selected WDs. Section~\ref{sec:4} presents our results on the cooling delay and implications for the current evolutionary models. We conclude and summarize the work in Section~\ref{sec:5}.

\begin{figure}
	\includegraphics[width=0.481\textwidth]{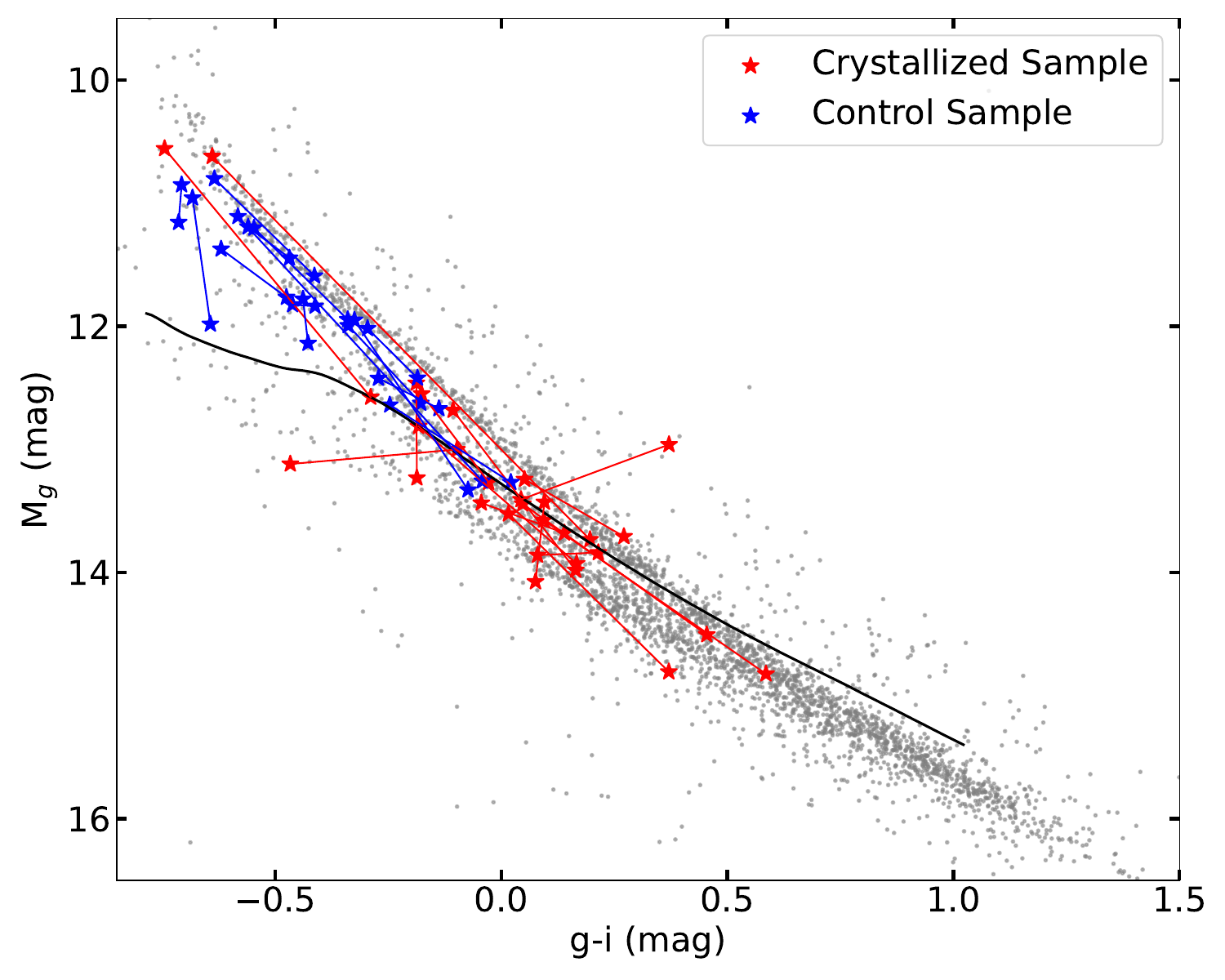}
    \caption{Color-magnitude diagram of the selected wide double white dwarf binaries using Pan-STARRS photometry. The red linked stars show our selected sample where one member is crystallized, the blue linked stars show the control sample where none of the members are crystallized, and the grey dots in the background are white dwarfs within 100 pc of the Sun \citep{kilic2020}. The black solid line shows the onset of crystallization based on \citep{bedard2020} models.}
   \label{fig:1}
\end{figure}

\vskip 8mm

\section{Wide Binary Sample Selection and Observations}
\label{sec:2}

\subsection{Sample Selection}

\citet{elbadry2021} identified around 1400 high-probability DWDs, 407 of which have photometry in the SDSS and Pan-STARRS for both components. Only 82 pairs in this sample have spectral types provided in the literature for both components. In order to identify wide binaries with potentially crystallized components, we use the photometric technique described in Section \ref{sec:3.1} below to constrain the mass and the crystallized core fraction of each WD assuming a pure hydrogen atmosphere (if no spectral information is available) and the C/O core evolutionary models from \citet{bedard2020}. We select binaries with a non-crystallized member and a companion that is $\geq1$\% crystallized. This reduces our sample to 108 candidate pairs. The crystallization fraction strongly depends on the internal composition and the mass of the WD. Given our assumption of pure H atmospheres for our initial sample selection, the crystallization fractions in this preliminary sample will be reevaluated later in Section~\ref{sec:3.2}.

\citet{heintz2022} showed that small uncertainties in WD mass (0.02-0.03 M$_{\odot}$) correspond to large uncertainties in the progenitor mass and main-sequence lifetimes for WDs with masses below 0.63 $M_{\odot}$. We further restrict our sample to WDs with $M\geq0.63~M_{\odot}$ to avoid this issue. Our preliminary crystallized sample comprises 18 DWDs, four of which have spectra available in the literature.

For comparison, we also select a control sample of DWDs where both members are not crystallized and both have masses above 0.63 M$_{\odot}$. This sample comprises 14 pairs, most of which were previously analyzed in the literature. 

\begin{figure*}
	\includegraphics[width=0.50\textwidth]{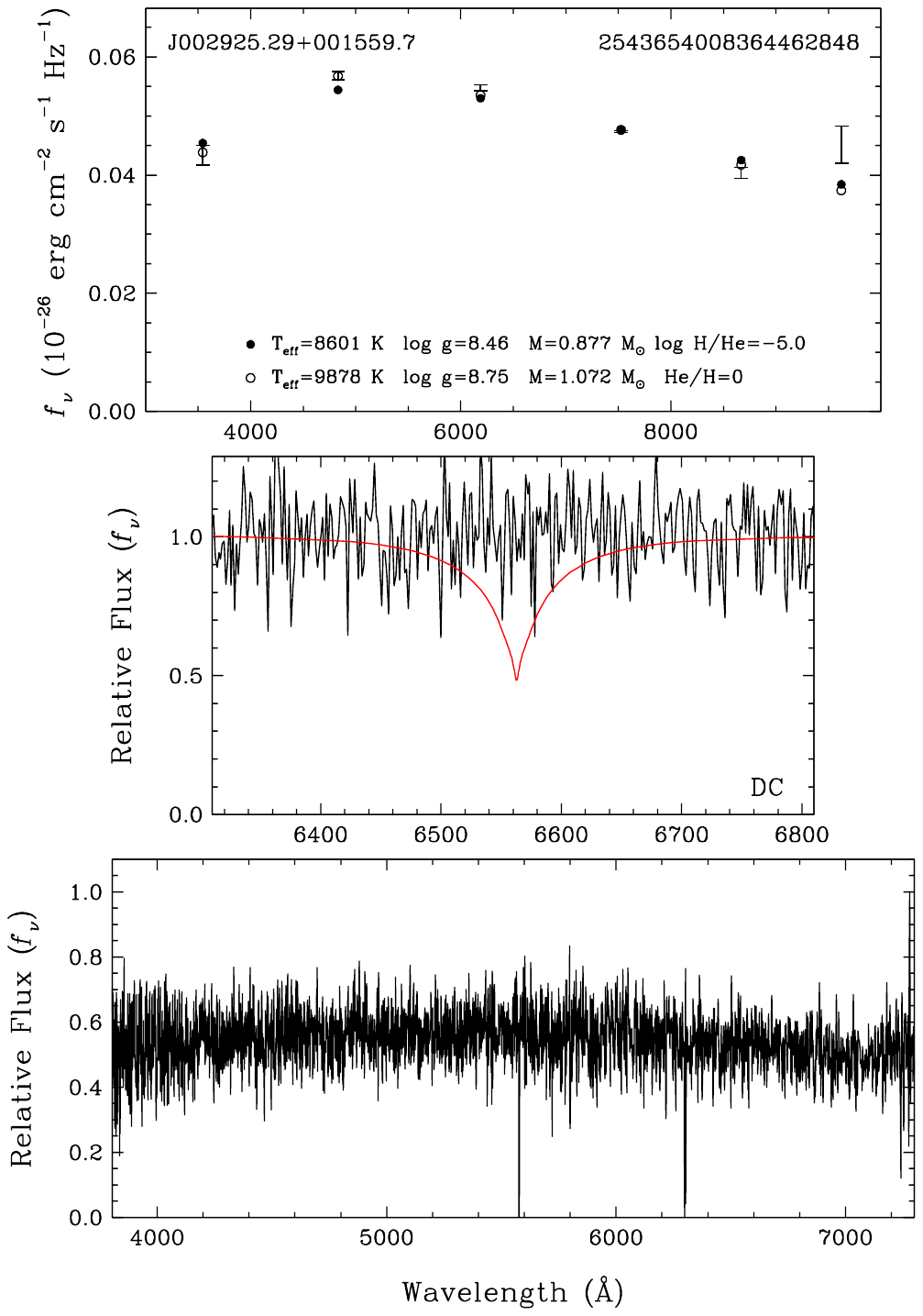}
        \includegraphics[width=0.50\textwidth]{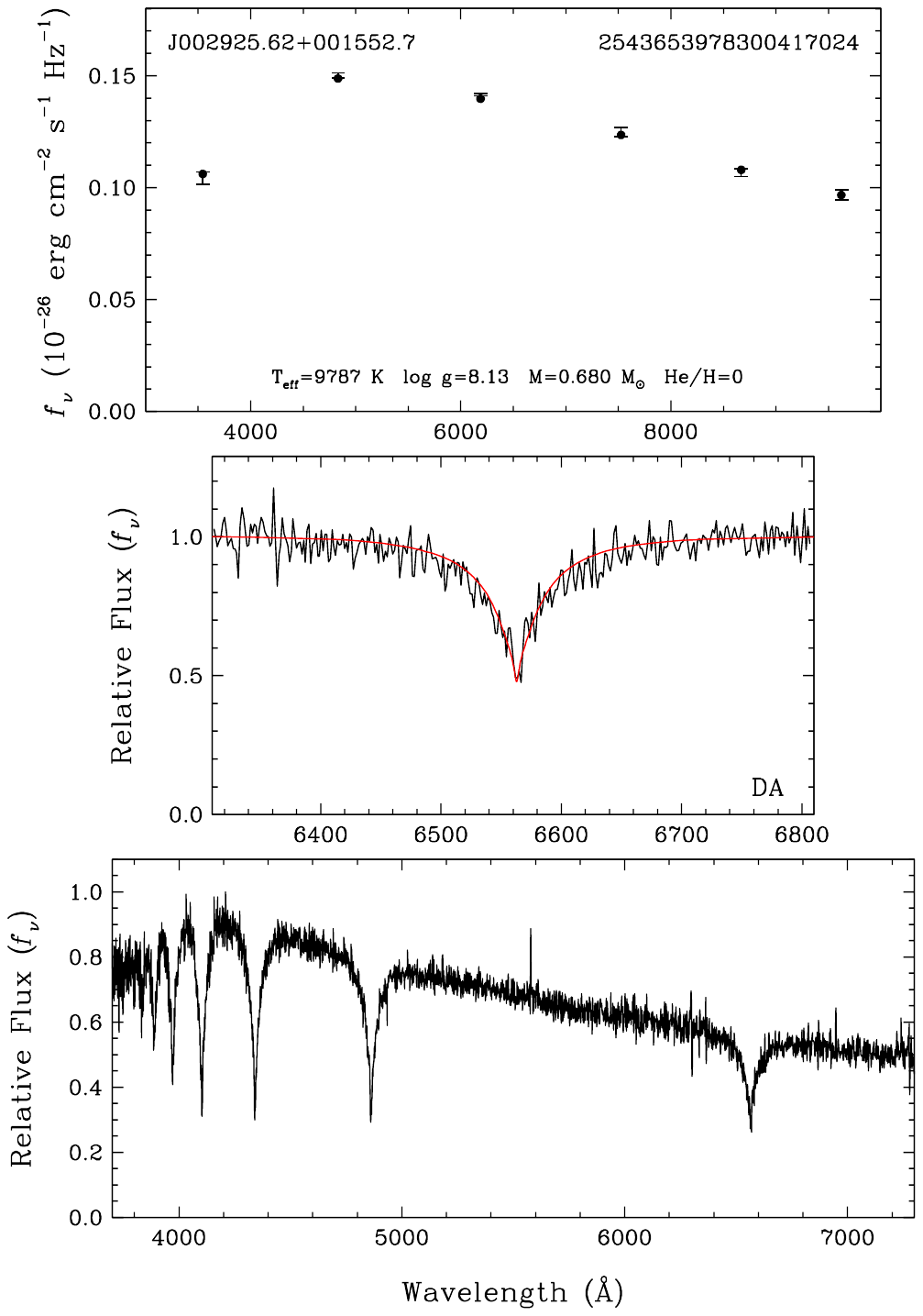}
    \caption{Example model fits for one of the pairs in our sample, including J002925.29+001559.7 (left panel) and J002925.62+001552.7 (right panel). The top panels show the best-fitting pure hydrogen (open circles on the left panel and filled circles on the right) and helium-rich (filled circles on the left panel) atmosphere white dwarf models to the SDSS-$u$ and PanSTARRS-$grizy$ photometry (error bars). The best-fitting parameters for each star are included in each panel. The fit for J002925.29+001559.7 favors the He-rich solution, and J002925.62+001552.7 favors a pure H atmosphere solution. The middle panels show the observed spectrum (black) in the H$\alpha$ region and the predicted H$\alpha$ (red) based on a pure H solution. The bottom panel shows the full spectral range for each observation, revealing additional H lines for J002925.62+001552.7 and no lines for J002925.29+001559.7. In this case, the WD on the left has a mass of 0.88 M$_{\odot}$ and a core that is 37\% crystallized, whereas the star on the right has a mass of 0.68 M$_{\odot}$ and 0\% crystallization of the core. This is one of the benchmark systems in this work. The complete figure set is available in the online version}
   \label{fig:2}
\end{figure*}

Figure~\ref{fig:1} shows the locations of our preliminary sample of crystallized systems (red linked stars) along with the control sample (blue linked stars) in a color-magnitude diagram (CMD) using the Pan-STARRS photometry. Grey dots in the background represent the 100 pc WD sample by \citet{kilic2020}, and the black solid line is the onset of crystallization based on the \citet{bedard2020} models. 
A few of the WDs in the preliminary crystallized sample appear slightly above the crystallization onset curve. However, classification based on a single color-magnitude diagram can be misleading, as the core-crystallization fraction strongly depends on the mass and effective temperature, which require precise constraints on the atmospheric composition. Hence, a detailed model atmosphere analysis of each system is necessary to constrain their physical parameters, as discussed below in Section~\ref{sec:3.2}.

\subsection{Observations}
\label{sec:2.1}

Out of the 18 systems in our preliminary selection, we targeted the 14 systems without optical spectroscopy in the literature for follow-up spectroscopy observations and succeeded in observing 11 of the pairs. We obtained follow-up optical spectroscopy of five pairs using the 8.1m Gemini North and South telescopes located in Cerro Pachón, Chile, and Mauna Kea, Hawaii, respectively. We were awarded time to observe three additional pairs at Gemini, but those observations could not be completed in the Gemini queue.

We used the Gemini Multi-Object Spectrograph (GMOS-N and GMOS-S) as part of the queue programs GN-2023A-Q-225, GN-2023A-Q-325, and GS-2023A-Q-325. We used the $1.0\arcsec$ slit and the B600 grating centered at 515 nm with GMOS-S, providing wavelength coverage from 3670 to 6800 Å and a resolution of 2.0 \text{\AA} per pixel in the 4×4 binned mode. We oriented the slit to match the binary position angle so that both components were observed at the same time. We used the same slit, central wavelength, and binning for our Gemini North observations but used the B480 grating, which provides improved sensitivity and wavelength coverage. 

We obtained follow-up optical spectroscopy of six pairs using the Blue Channel Spectrograph \citep{schmidt1989} on the 6.5m MMT. We used the 500 lines mm$^{-1}$ grating and the $1.25\arcsec$ slit, which provides a wavelength coverage of 3800 to 6900 \text{\AA} and a spectral resolution of 4.7 \AA. We reduced the data from both telescopes using the standard IRAF routines. In particular, we used the GMOS package in IRAF for the Gemini-GMOS data. All of our data have Signal-to-Noise (S/N) above 15 per pixel, enabling accurate spectral classification of each system.

\begin{figure*}
\centering
	\includegraphics[width=\textwidth]{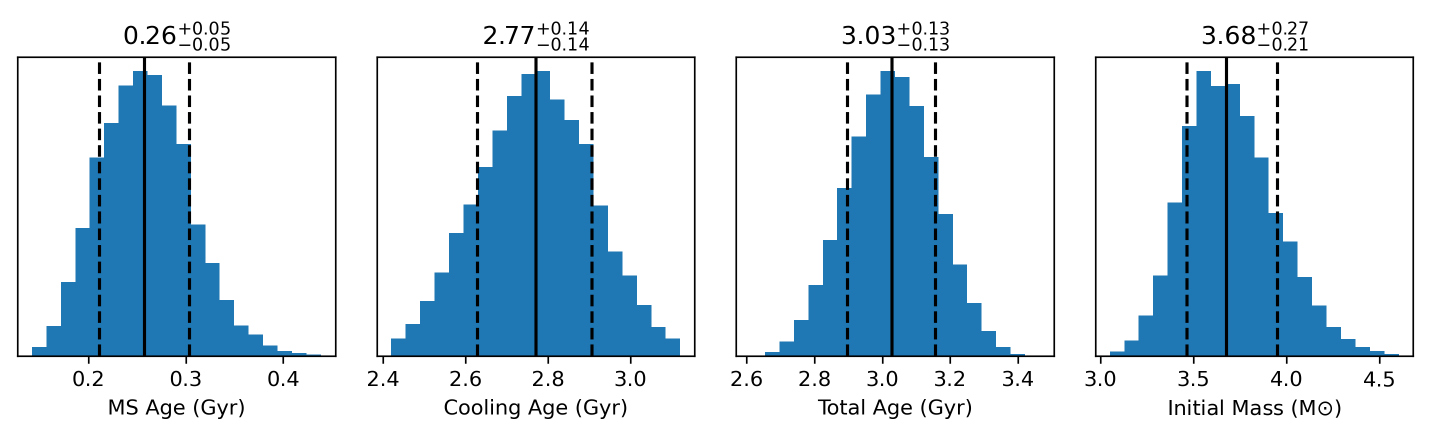}
        \includegraphics[width=\textwidth]{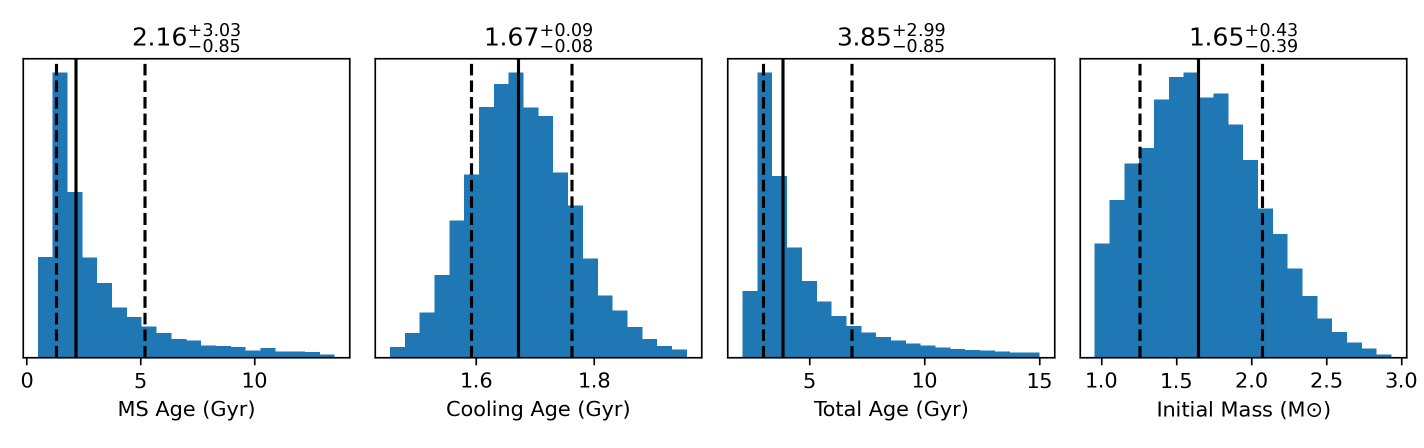}
    \caption{Probability distribution function output of J1113+3238 pair from \texttt{wdwarfdate} by \citet{kiman2022} for the main sequence age (or progenitor's lifetime), cooling age, total age of the WD, and the initial mass determinations. The calculated values are shown in each panel. In particular, we used \citet{bauer2023} models to obtain the white dwarf cooling age. The top panels show the parameters for J111322.48+323859.0, a 0.88 M$_{\odot}$ WD with a $\sim$5\% precision in its total age, whereas the bottom panels show the parameters for a 0.65 M$_{\odot}$ WD with significantly larger errors in its total age estimate.}
   \label{fig:3}
\end{figure*}

\subsection{Literature Spectra}
\label{sec:2.2}

For the control sample of 14 pairs where both components are more massive than 0.63 $M_{\odot}$ and are not crystallized, we found optical spectra for six pairs in \citet{andrews2015}, which were observed at the Apache Point Observatory 3.5m telescope.  We found spectra for six additional pairs in the Montreal White Dwarf Database (MWDD)\footnote{\url{https://www.montrealwhitedwarfdatabase.org/}} \citep{dufour2017}. For one of the pairs, J0859$+$4250 binary, we could not find an optical spectrum and instead relied on the spectral classification in the literature. For another pair, we found a spectrum for one of the components, J222236.30$-$082807.9, but not the other member. 

\section{Analysis}
\label{sec:3} 

\subsection{Model Atmosphere Analysis}
\label{sec:3.1}

To get an accurate estimation of the physical properties of our WD sample, we use the photometric technique described in \citet{bergeron2019}. We apply this approach to our analysis using the SDSS $u$ and Panoramic Survey Telescope and Rapid Response System (Pan-STARRS, \citet{panstarrs2016}) $grizy$ photometry. Briefly, this method involves transforming the observed magnitudes in the different bandpasses into average fluxes by using the correct zero points. Then, using appropriate model atmospheres and chemical compositions, synthetic fluxes are generated and compared to these observed fluxes via $\chi^2$ minimization to derive the best-fitting parameters and their uncertainties. We fit for the effective temperature and the solid angle, $\pi (R/D)^2$, where $D$ is the distance to the star, and $R$ is the radius. Since distances are known from {\emph Gaia} parallaxes \citep{bj2021}, we can constrain the radius of the star directly and, therefore, the gravity (logg parameter) and the mass using \citet{bedard2020} evolutionary models. Considering the distances of our sample (d$>$100 pc), we also correct for reddening by using the \texttt{STILISM}\footnote{\url{https://stilism.obspm.fr/}} values \citep{stilism2014,stilism2017}. The details of our fitting method are further discussed in \citet{kilic2020}. 

Figure~\ref{fig:2} shows the results from this analysis for one of the pairs. For each star, the top panel shows the SDSS $u$ and Pan-STARRS $grizy$ photometry (error bars), along with the predicted fluxes from the best-fitting model (filled and open circles). The labels in the same panel give each source's {\emph Gaia} Source ID, object name, and best-fitting parameters. The middle panel shows the predicted spectrum (red line) based on the pure hydrogen solution, along with the observed spectrum in the H$\alpha$ region, and the bottom panel shows the entire spectrum for each star.

The brighter object in the first pair shown in Figure~\ref{fig:2}, J002925.62+001552.7 (right panel), is a DA WD that shows Balmer lines, and the best-fitting pure H atmosphere model has $T_{\rm eff} = 9787 \pm 148$ K. The fainter companion, J002925.29+001559.7 (left panel), is a DC WD with no visible absorption features. If this were a pure H atmosphere, we would have seen a relatively strong H$\alpha$ line, the absence of which indicates that the atmosphere is dominated by helium. Here, the best-fitting helium-rich model has $T_{\rm eff} = 8601 \pm 156$ K, and the abundance ratio of H to He is given by log H/He= $-5.0$. The evolutionary models predict that the former has a mass M= 0.68 $\pm$ 0.04 M$_{\odot}$ and is not crystallized, whereas the latter has a mass M= 0.88 $\pm$ 0.04 M$_{\odot}$ and its core is $\sim$37\% crystallized. This is one of the benchmark systems used in this work to constrain the cooling delay in WDs. The complete figure set of model fits is available in the online version of this article.

Our preliminary crystallized sample comprises eight DA-DA, three DA-DQ, two DAH-DA pairs, one DA-DC, and another DAH-DC pair. On the other hand, the control sample comprises 13 DA-DA pairs and one DA-DC pair. For the three DQs in our sample, we rely on the models by \citet{blouin2019}. In this case, we use neutral C I lines or the C$_2$ Swan band to fit for C/He. We fit both the photometric and spectroscopic data in an iterative process until we converge to a consistent solution. For the magnetic WDs in our sample, we use a similar approach to \citet{moss2024}, where the total line opacity is calculated as the sum of the individual Stark-broadened Zeeman components \citep[see][for further details]{moss2024}.  

Table~\ref{tab:1} and Table~\ref{tab:A1} present the best-fitting physical parameters for our preliminary crystallized sample and control sample, respectively, including $T_{\rm eff}$, $\log{g}$, and mass.

\begin{deluxetable*}{crrrrrrrccc}[htp]
\tabletypesize{\footnotesize}
\tablecolumns{11} \tablewidth{0pt}
\tablecaption{Selected Wide Double White Dwarfs with a Crystallized Member} 
\label{tab:1}
\tablehead{\colhead{Object name} & \colhead{{\emph Gaia} Source ID} & \colhead{Type} & \colhead{T$_{\text{eff}}$} & \colhead{log(g)} & \colhead{M$_{\text{WD}}$ } & \colhead{$\tau_{\text{cooling}}$ } & \colhead{$\tau_{\text{progenitor}}$} & \colhead{Total Age} & \colhead{$\Delta$ Age$^a$} & \colhead{Crystallized Fraction$^{b}$} \\
& \colhead{(DR3)} & & \colhead{(K)}  & \colhead{(cm s$^{-1}$)} & \colhead{(M$_{\odot}$)} &  \colhead{(Gyr)} & \colhead{(Gyr)} & \colhead{(Gyr)} & \colhead{(Gyr)} & \colhead{($M/M_{\star}$)} }
\rotate
\startdata
\hline\hline
J002925.29+001559.7 & 2543654008364462848 & DC & 8601 $\pm$ 156 & 8.46 $\pm$ 0.04 & 0.88 $\pm$ 0.04 & 2.10$^{+0.21}_{-0.21}$ & 0.24$^{+0.06}_{-0.05}$ & 2.34$^{+0.19}_{-0.18}$ & \multirow{2}{*}{-0.21$^{+1.57}_{-0.57}$} & \multirow{2}{*}{0.37(0.15)}\\
J002925.62+001552.7 & 2543653978300417024 & DA & 9787 $\pm$ 148 & 8.13 $\pm$ 0.04 & 0.68 $\pm$ 0.04 & 0.71$^{+0.06}_{-0.05}$ & 1.39$^{+1.62}_{-0.56}$ & 2.11$^{+1.57}_{-0.54}$ & &\\
\hline
J081427.92+013325.6 & 3089916403229120640 & DA & 9481 $\pm$ 414 & 8.47 $\pm$ 0.14 & 0.90 $\pm$ 0.12 & 1.56$^{+0.56}_{-0.50}$ & 0.25$^{+0.20}_{-0.09}$ & 1.88$^{+0.49}_{-0.33}$ & \multirow{2}{*}{1.21$^{+4.99}_{-1.04}$} & \multirow{2}{*}{ 0.12(0.00)}\\
J081427.83+013318.7 & 3089916398933726592 & DA & 8281 $\pm$ 201 & 8.15 $\pm$ 0.14 & 0.69 $\pm$ 0.12 & 1.04$^{+0.23}_{-0.14}$ & 1.96$^{+5.04}_{-1.28}$ & 3.00$^{+5.00}_{-0.95}$ & &\\
\hline
J102142.06+394225.4 & 804040486519166976 & DA & 7692 $\pm$ 303 & 8.29 $\pm$ 0.13 & 0.78 $\pm$ 0.12 & 1.61$^{+0.79}_{-0.37}$ & 0.71$^{+1.60}_{-0.38}$ & 2.69$^{+1.05}_{-0.45}$ & \multirow{2}{*}{0.97$^{+5.00}_{-1.29}$} & \multirow{2}{*}{ 0.18(0.00)}\\
J102141.29+394215.5 & 804040108562044288 & DA & 7846 $\pm$ 203 & 8.13 $\pm$ 0.12 & 0.67 $\pm$ 0.11 & 1.19$^{+0.23}_{-0.15}$ & 2.15$^{+5.08}_{-1.34}$ & 3.33$^{+5.00}_{-1.05}$ & &\\
\hline
J105242.54+283252.0 & 731411283875749760 & DA & 6171 $\pm$ 98 & 8.24 $\pm$ 0.05 & 0.74 $\pm$ 0.04 & 3.18$^{+0.34}_{-0.32}$ & 0.69$^{+0.52}_{-0.22}$ & 3.97$^{+0.35}_{-0.30}$ & \multirow{2}{*}{4.23$^{+4.35}_{-3.57}$} & \multirow{2}{*}{ 0.41(0.23)}\\
J105242.57+283255.3 & 731411283874346240 & DQ & 7061 $\pm$ 77 & 7.95 $\pm$ 0.05 & 0.55 $\pm$ 0.05 & 1.31$^{+0.10}_{-0.09}$ & 6.84$^{+4.32}_{-3.58}$ & 8.18$^{+4.34}_{-3.56}$ & & \\
\hline
J111322.48+323859.0 & 757911988004305280 & DA & 7661 $\pm$ 75 & 8.45 $\pm$ 0.02 & 0.88 $\pm$ 0.02 & 2.77$^{+0.14}_{-0.14}$ & 0.26$^{+0.05}_{-0.05}$ & 3.03$^{+0.13}_{-0.13}$ & \multirow{2}{*}{0.82$^{+2.99}_{-0.86}$} & \multirow{2}{*}{0.52(0.40)}\\
J111319.38+323818.0 & 757911884925087104 & DA & 6882 $\pm$ 77 & 8.09 $\pm$ 0.03 & 0.65 $\pm$ 0.03 & 1.67$^{+0.09}_{-0.08}$ & 2.16$^{+3.03}_{-0.85}$ & 3.85$^{+2.99}_{-0.85}$ & &\\
\hline
J115749.13+313931.0 & 4026615235380699392 & DAH & 8734 $\pm$ 202 & 8.44 $\pm$ 0.06 & 0.88 $\pm$ 0.05 & 1.91$^{+0.30}_{-0.30}$ & 0.26$^{+0.07}_{-0.06}$ & 2.17$^{+0.27}_{-0.25}$ & \multirow{2}{*}{2.03$^{+4.51}_{-1.29}$} & \multirow{2}{*}{ 0.28(0.01)}\\
J115749.39+313931.0 & 4026615235380699520 &DA & 6996 $\pm$ 133 & 8.08 $\pm$ 0.07 & 0.64 $\pm$ 0.06 & 1.56$^{+0.19}_{-0.15}$ & 2.60$^{+4.58}_{-1.35}$ & 4.18$^{+4.51}_{-1.26}$ & & \\
\hline
J115937.82+134408.7 & 3920275276810355200 & DA & 14833 $\pm$ 490 & 9.11 $\pm$ 0.02 & 1.19 $\pm$ 0.01 & 1.46$^{+0.07}_{-0.07}$ & 0.06$^{+0.01}_{-0.01}$ & 1.51$^{+0.08}_{-0.07}$ & \multirow{2}{*}{0.18$^{+0.22}_{-0.13}$} & \multirow{2}{*}{0.88(0.77)}\\
J115937.81+134413.9 & 3920275276810355072 & DA & 8998 $\pm$ 90 & 8.26 $\pm$ 0.02 & 0.76 $\pm$ 0.02 & 1.14$^{+0.05}_{-0.05}$ & 0.53$^{+0.23}_{-0.10}$ & 1.68$^{+0.21}_{-0.10}$ &  &\\
\hline
J123647.95+682501.6 & 1682366418152856448 & DA& 6863 $\pm$ 116 & 8.17 $\pm$ 0.04 & 0.70 $\pm$ 0.03 & 1.95$^{+0.22}_{-0.17}$ & 1.10$^{+0.75}_{-0.43}$ & 3.10$^{+0.66}_{-0.39}$ & \multirow{2}{*}{3.57$^{+4.70}_{-3.16}$} & \multirow{2}{*}{ 0.10(0.003)}\\
J123647.42+682502.9 & 1682554091043762560 & DQ & 8813 $\pm$ 133 & 7.99 $\pm$ 0.03 & 0.57 $\pm$ 0.03 & 0.77$^{+0.05}_{-0.05}$ & 5.79$^{+4.71}_{-3.13}$ & 6.55$^{+4.68}_{-3.11}$ & & \\
\hline
J134739.13+251228.9 & 1444442547261998464 & DA & 6743 $\pm$ 104 & 8.13 $\pm$ 0.06 & 0.67 $\pm$ 0.05 & 1.88$^{+0.24}_{-0.19}$ & 1.54$^{+2.65}_{-0.67}$ & 3.50$^{+2.46}_{-0.56}$ & \multirow{2}{*}{0.02$^{+3.23}_{-1.38}$} & \multirow{2}{*}{ 0.08(0.00)}\\
J134737.58+251233.1 & 1444442512902260864 & DA & 8194 $\pm$ 133 & 8.13 $\pm$ 0.05 & 0.67 $\pm$ 0.04 & 1.10$^{+0.10}_{-0.09}$ & 1.58$^{+2.71}_{-0.69}$ & 2.70$^{+2.63}_{-0.63}$ & &\\
\hline
J135834.36+263345.2 & 1450779342012324224 & DA& 6717 $\pm$ 195 & 8.29 $\pm$ 0.06 & 0.78 $\pm$ 0.05 & 2.78$^{+0.43}_{-0.43}$ & 0.50$^{+0.35}_{-0.13}$ & 3.38$^{+0.36}_{-0.34}$ & \multirow{2}{*}{-1.31$^{+2.64}_{-0.82}$} & \multirow{2}{*}{ 0.45(0.11)}\\
J135834.66+263343.1 & 1450779346306149760 & DA& 21694 $\pm$ 453 & 8.04 $\pm$ 0.03 & 0.65 $\pm$ 0.02 & 0.06$^{+0.01}_{-0.01}$ & 1.97$^{+2.64}_{-0.73}$ & 2.02$^{+2.64}_{-0.73}$ & \\
\hline
J185722.65+781332.1 & 2293210651405001216 &DA & 7709 $\pm$ 755 & 8.25 $\pm$ 0.21 & 0.75 $\pm$ 0.19 & 1.57$^{+1.08}_{-0.48}$ & 1.39$^{+4.64}_{-0.99}$ & 3.48$^{+3.98}_{-1.04}$ & \multirow{2}{*}{-0.38$^{+4.10}_{-1.49}$} & \multirow{2}{*}{0.07(0.00)}\\
J185722.00+781332.2 & 2293210651402924160 &DA & 8041 $\pm$ 120 & 8.22 $\pm$ 0.10 & 0.73 $\pm$ 0.10 & 1.26$^{+0.37}_{-0.17}$ & 1.03$^{+2.02}_{-0.57}$ & 2.41$^{+1.77}_{-0.39}$ & &\\
\hline
J211658.03+082047.6 & 1740077893712777216 & DA & 6302 $\pm$ 244 & 8.15 $\pm$ 0.09 & 0.68 $\pm$ 0.08 & 2.77$^{+0.86}_{-0.61}$ & 1.41$^{+2.86}_{-0.72}$ & 4.40$^{+2.88}_{-1.05}$ & \multirow{2}{*}{-} & \multirow{2}{*}{0.22(0.00)} \\
J211657.92+082048.8$^{c}$ & 1740077893710129536 & DQ & 7039 $\pm$ 90 & 7.86 $\pm$ 0.06 & 0.49 $\pm$ 0.05 & 1.33$^{+0.12}_{-0.12}$ & - & - & & \\
\hline
J213102.82+083203.8 & 1741031891851063424 & DA& 7065 $\pm$ 88 & 8.26 $\pm$ 0.02 & 0.75 $\pm$ 0.03 & 2.28$^{+0.17}_{-0.17}$ & 0.56$^{+0.26}_{-0.13}$ & 2.88$^{+0.23}_{-0.19}$ & \multirow{2}{*}{0.61$^{+3.51}_{-1.26}$} & \multirow{2}{*}{0.25(0.09)}\\
J213103.09+083202.6 & 1741031896140411648 & DA& 9048 $\pm$ 52 & 8.06 $\pm$ 0.02 & 0.64 $\pm$ 0.02 & 0.80$^{+0.02}_{-0.02}$ & 2.67$^{+3.49}_{-1.24}$ & 3.47$^{+3.51}_{-1.24}$ & & \\
\hline
J225932.73+140444.2 & 2815944352131513472 & DAH & 10768 $\pm$ 159 & 8.49 $\pm$ 0.02 & 0.91 $\pm$ 0.02 & 1.10$^{+0.08}_{-0.06}$ & 0.22$^{+0.05}_{-0.04}$ & 1.32$^{+0.08}_{-0.07}$ & \multirow{2}{*}{-0.96$^{+0.09}_{-0.08}$} & \multirow{2}{*}{ 0.02(0.00)}\\
J225932.21+140439.2 & 2815944352131513088 & DA& 27799 $\pm$ 402 & 8.33 $\pm$ 0.02 & 0.84 $\pm$ 0.02 & 0.03$^{+0.00}_{-0.00}$ & 0.33$^{+0.06}_{-0.05}$ & 0.36$^{+0.06}_{-0.05}$ & \\
\hline
J231939.16-035857.8 & 2634002940402436480 & DAH & 5652 $\pm$ 52 & 8.09 $\pm$ 0.03 & 0.64 $\pm$ 0.03 & 3.13$^{+0.24}_{-0.26}$ & 2.48$^{+3.32}_{-1.11}$ & 5.60$^{+3.26}_{-1.03}$ & \multirow{2}{*}{4.65$^{+4.52}_{-4.01}$} & \multirow{2}{*}{ 0.31(0.06)}\\
J231938.65-035833.1 & 2634003146560869248 & DC & 6920 $\pm$ 80 & 7.92 $\pm$ 0.03 & 0.53 $\pm$ 0.02 & 1.29$^{+0.07}_{-0.06}$ & 8.34$^{+3.58}_{-3.70}$ & 9.60$^{+3.57}_{-3.67}$ & \\
\hline\hline
\enddata
\tablenotetext{\tiny a}{$\Delta$Age is the difference in age between the crystallized component and the non-crystallized companion following the \citet{barlow2003} and \citet{laursen2019} prescription for asymmetric errors.}
\tablenotetext{\tiny b}{Crystallized core fraction for the crystallized component and its 3$\sigma$ lower limit in parenthesis calculated using \citet{bauer2023} models.}
\tablenotetext{\tiny c}{The mass of J211657.92+082048.8 is too small to determine a progenitor mass. Therefore, no age estimation can be made.} 
\end{deluxetable*}

\subsection{White Dwarf Age Determination}
\label{sec:3.2}

In order to obtain the total age of a WD, we need to determine both the cooling age ($\tau_{\text{cool}}$) and the progenitor lifetime ($\tau_{\text{prog}}$). The former is the time since the star ended up on the WD cooling track, and the latter is the time from the zero-age main sequence (ZAMS) to the first thermal pulse of the Asymptotic Giant Branch \citep[e.g.,][]{barrientos2021,heintz2022}. The best-fitting model parameters and the evolutionary models enable us to constrain the cooling age of each WD. However, determining the progenitor lifetime is more complex as we have to rely on empirically calibrated IFMR to trace back the star's initial mass on the main sequence and calculate its lifetime. 

To calculate the progenitor mass of each WD, we take advantage of the Bayesian method implemented in \texttt{wdwarfdate}\footnote{\url{https://github.com/rkiman/wdwarfdate}} by \citet{kiman2022}. We use the best-fitting effective temperature and logg from our model atmosphere analysis described in Section~\ref{sec:3.1} along with the C/O core evolutionary models by \citet{bauer2023} and O/Ne core cooling sequences by \citet{camisassa2019} to calculate the cooling age, $\tau_{\text{cool}}$. We also use the \texttt{MIST} based IFMR from \citet{cummings2018}, and MIST evolutionary tracks from \citet{choi2016} to estimate the initial mass, and the $\tau_{\text{prog}}$. Since the tangential velocities of our sample are consistent with the thin disk and solar neighborhood kinematics \citep{chiba2000}, we use [Fe/H]=0.0 and no rotation (v/v$_{\text{crit}}$=0). The total age of the WD is obtained by adding $\tau_{\text{cool}}$ and $\tau_{\text{prog}}$ (see \citealt{kiman2022} for potential caveats). A major source of uncertainty in the WD age determination is the $\tau_{\text{prog}}$ calculation. We discuss the sensitivity of our results to the assumed IFMR and the progenitor star metallicity in Section~\ref{sec:4.2}. The cooling ages, progenitor lifetimes, and total ages for the preliminary crystallized sample and the control sample are presented in Table~\ref{tab:1} and Table~\ref{tab:A1}, respectively.

One of the WDs in our sample, J211657.92+082048.8, has a mass of 0.49 M$_{\odot}$, which indicates that it has likely formed through close binary evolution. Hence, this binary was likely a triple system in the past \citep[see e.g.,][]{andrews2016,coutu2019}. Considering that we cannot constrain the age of this system, we remove it from further consideration, reducing the number of crystallized systems to 14.

An illustration of the \texttt{wdwarfdate} output is shown in Figure~\ref{fig:3} for J1113+3238 system. Both panels (top and bottom) show the probability distribution function for the main sequence age, cooling age, total age, and the initial mass of each binary component. The top panel shows J111322.48+323859.0, a 0.88 M$_{\odot}$ WD with a $\sim$5\% uncertainty in its total age. For comparison, the bottom panel shows J111319.38+323818.0, a 0.65 M$_{\odot}$ WD with a much more uncertain total age. 

We provide a comparison between our results and the parameters obtained by \citet{heintz2022} in Table~\ref{tab:A2} and Figure~\ref{fig:A1} in the Appendix. \citet{heintz2022} assumed a pure H atmosphere for their targets. The ages from our analysis and \citet{heintz2022} agree within the errors for the majority of our targets, though we find three significant outliers that are non-DA white dwarfs (DC and DQ spectral types). Since our analysis takes advantage of follow-up spectroscopy of all targets and the model atmospheres with appropriate chemical composition, the physical parameters in this study are more reliable.

Two of the DWDs in our control sample are also included in \citet{holland2024}. For the J1313+2030 binary, we measure total ages of 0.88$^{+0.06}_{-0.05}$ and 1.19$^{+0.43}_{-0.22}$ Gyr for the two components, whereas \citet{holland2024} estimated 0.79$^{+0.08}_{-0.05}$ and 0.76$^{+0.16}_{-0.08}$ Gyr for the same stars, respectively. Similarly, for the J2223+2201 binary, we estimated 0.44$^{+0.04}_{-0.04}$ and 0.91$^{+0.20}_{-0.09}$ Gyr, while they estimated 0.35$^{+0.08}_{-0.06}$ and 0.79$^{+0.17}_{-0.12}$ Gyr, respectively. These estimates are consistent with our results within the errors. 

We constrain the crystallized core fractions using the C/O core evolutionary models from \citet{bauer2023} for WDs with M$\leq$ 1.03 M$_{\odot}$, and O/Ne core models from \citet{camisassa2019} for WDs with M$\geq$ 1.10 M$_\odot$. Table~\ref{tab:1} includes the crystallized core fraction and its 3$\sigma$ lower limit in parenthesis. We exclude the pairs where the 3$\sigma$ lower limit indicates a non-crystallized core; specifically, this cut removes J0814+0133, J1021+3942, J1347+2512, J1857+7813, and J2259+1404 binaries from further consideration, reducing the number of significantly crystallized pairs to nine. 

\vskip 8mm
\section{Discussion}
\label{sec:4}

\subsection{Quantifying the Cooling Delays in C/O White Dwarfs}
\label{sec:4.1}

For the nine DWDs with a crystallized member, we estimate the total ages for each WD using the approach explained in detail in Section~\ref{sec:3.2}. In order to test the age estimates from the current evolutionary models that include the delays from the latent heat and C/O phase separation of crystallization \citep[see][]{bauer2023}, we calculate the difference in total age, $\Delta$Age, by subtracting the age of the crystallized member from the age of the non-crystallized member. If there are cooling delays that are unaccounted for in the current evolutionary models, then we expect the age of the crystallized member to be underestimated; hence, we expect a positive $\Delta$Age. Since the ages have asymmetric errors and non-Gaussian probability distribution functions (see Figure~\ref{fig:3}), we used the approach described in \citet{barlow2003} and implemented by \citet{laursen2019} (see their Appendix B) to subtract these quantities\footnote{We used the python package \texttt{add\_asym} (\url{https://github.com/anisotropela/add_asym}).}. For our sample, the crystallized core mass fraction ranges from 10\% to 88\%, and the masses range from 0.63 M$_{\odot}$ to 1.21 M$_{\odot}$, covering a significant portion of the parameter space for WDs. 

We use the J1113+3238 binary to demonstrate our methodology. This binary consists of a $M=0.86 \pm 0.01~M_{\odot}$ WD with a 52\% crystallized core and a non-crystallized $M=0.63 \pm 0.01~M_{\odot}$ companion. The cooling age for the crystallized member is 2.77$^{+0.14}_{-0.14}$ Gyr. With an estimated progenitor lifetime of 0.26$^{+0.05}_{-0.05}$ Gyr, the total age is constrained to 3.03$^{+0.13}_{-0.13}$ Gyr. On the other hand, the non-crystallized member has a cooling age of 1.67$^{+0.09}_{-0.08}$ Gyr, progenitor lifetime of 2.16$^{+3.03}_{-0.85}$ Gyr, and a total age estimate of 3.85$^{+2.99}_{-0.85}$ Gyr. Hence, the evolutionary models underpredict the age of the crystallized member by $\Delta$Age = 0.82$^{+2.99}_{-0.86}$ Gyr. 

\begin{figure*}
\centering
	\includegraphics[width=\textwidth]{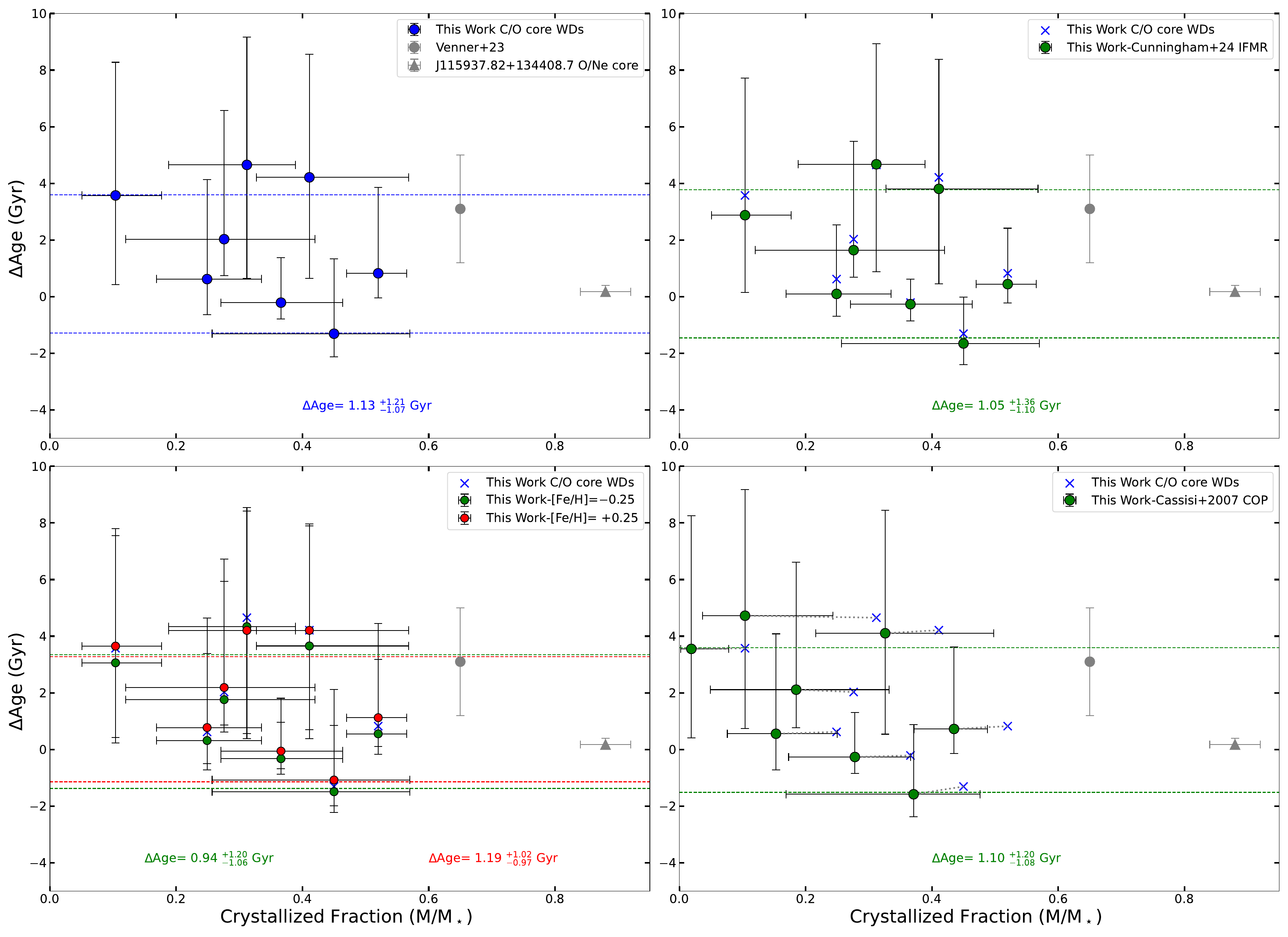}
    \caption{Cooling anomaly in wide double white dwarfs as a function of the crystallized core fraction. The top left panel shows the main results of this work using the prescription described in Section~\ref{sec:3.2}. The blue dots represent the 8 wide double white dwarfs with C/O cores used in this work. The gray dot is the benchmark obtained by \citet{venner2023}, in agreement with our results within 1$\sigma$. The gray triangle is an O/Ne white dwarf in our sample that is not part of the analysis. The blue dashed lines show the upper and lower 3$\sigma$ level (99.7\%). We found a weighted mean $\Delta$Age = 1.13$^{+1.21}_{-1.07}$ Gyr. The top right panel shows the sensitivity of our results to the initial-final mass relation (IFMR). The blue crosses are our default results (same as the top left panel), the green dots show the $\Delta$Age using the \citet{cunningham2024} IFMR, and the green dashed line shows upper and lower 3$\sigma$ levels. In this case, the calculated weighted mean is $\Delta$Age = 1.05$^{+1.36}_{-1.10}$ Gyr. The bottom left panel shows the sensitivity of our results to the progenitor star metallicity. Our default results use solar metallicity [Fe/H]=0 (blue crosses), whereas green and red dots show the $\Delta$Age values for [Fe/H]=$-$0.25 and [Fe/H]=$+$0.25, respectively. The green and red lines show the 3$\sigma$ limits. $\Delta$Age ranges from 0.94$^{+1.20}_{-1.06}$ Gyr to 1.19$^{+1.02}_{-0.97}$ Gyr for these metallicities. The bottom right panel shows the sensitivity of our results to the conductive opacities used in the cooling models. Our default model (blue crosses) is based on the \citet{bauer2023} cooling models with electron conductive opacities from \citet{blouin2020b}. The green dots show the results when using electron conductive opacities from \citet{cassisi2007}. The green dashed lines show the 3$\sigma$ limits and the gray dotted lines show the change from the default results. In this case, $\Delta$Age = 1.10$^{+1.20}_{-1.08}$ Gyr.}
   \label{fig:4}
\end{figure*}

The top left panel of Figure~\ref{fig:4} shows the main results of this paper: $\Delta$Age versus the core crystallized mass fraction for our sample of C/O core DWDs (blue dots). The dashed lines show the upper and lower 3$\sigma$ limits for the sample. Because of the asymmetric errors in the $\Delta$Age measurements, we estimate the weighted mean of the sample by bootstrapping 10,000 times and calculate a weighted mean value using the \citet{barlow2003} formulation (see their Section 7). After that, we select 50\% of the distribution as the mean, 16\% to 84\% as the 1$\sigma$ uncertainties, and 0.3\% to 99.7\% as the 3$\sigma$ limits. We find a cooling anomaly of $\Delta$Age = 1.13$^{+1.21}_{-1.07}$ Gyr from the 8 pairs with a crystallized C/O core member. For comparison, we also plot the Sirius-like benchmark system from \citet{venner2023}, where the cooling anomaly is +3.1$\pm$1.9 Gyr (gray dot), and J1159+1344, the only massive WD with a likely O/Ne core in our sample (gray triangle). The latter was not included in our $\Delta$Age estimate (see Section~\ref{sec:4.4}).

We perform a similar analysis on the 14 binaries in the control sample and find a weighted mean of $\Delta$Age = $-0.03 \pm 0.15$ Gyr (see Table~\ref{tab:A1} for details). This indicates that our method is reliable and that the evolutionary models give consistent results for non-crystallized WDs. Note that these models include the latent heat and the C/O phase separation upon crystallization \citep{bauer2023}, but not the extra energy from $^{22}$Ne distillation. Our sample of DWDs with crystallized members do not show a significant cooling delay, as the $\Delta$Age value is consistent with the null result within the errors. More importantly, we exclude cooling delays from $^{22}$Ne distillation and other neutron-rich impurities exceeding 3.6 Gyr for 0.6-0.9 $M_{\odot}$ WDs at the 99.7\% (3$\sigma$) confidence level. Even though one may expect a correlation between $\Delta$Age and the crystallized core fraction, we do not observe a significant trend in our sample. \citet{blouin2021a} suggested that the cooling delay from $^{22}$Ne distillation may kick in only after the core is 60\% crystallized. Unfortunately, all of the C/O core WDs in our DWD sample have core-crystallization fractions below this limit.

\subsection{Sensitivity to the Initial-Final Mass Relation, Metallicity, and Electron Conductive Opacity}
\label{sec:4.2}

The ultimate challenge in measuring WD ages is the determination of the progenitor's lifetime. To obtain this number, the use of an IFMR to trace back the initial mass and the use of evolutionary tracks to estimate the main sequence age are essential. The latter also heavily depends on the metallicity. 

To test the sensitivity of the total ages to the IFMR, we perform new calculations using the IFMR from \citet{cunningham2024} for a fixed metallicity [Fe/H]=0. The top right panel of Figure~\ref{fig:4} shows the results using this prescription. The green dots display the $\Delta$Age values for our crystallized sample, blue crosses show the results from the default prescription described in Section~\ref{sec:3.2}, and the green dashed line shows the upper and lower 3$\sigma$ limits. The average age discrepancy between both sets is $\sim$0.36 Gyr. This translates into $\Delta$Age values shifting slightly towards smaller values with a weighted average of $\Delta$Age = 1.05$^{+1.36}_{-1.10}$ Gyr. Therefore, the IFMR choice minimally impacts our results.

To test the impact of the progenitor star metallicity on our results, we used two more sets of MIST stellar evolution tracks with [Fe/H]=$-0.25$ and $+0.25$ \citep[see][]{rebassa2021} and no rotation (v/v$_{crit}=0$) for a fixed IFMR \citep{cummings2018}. Figure~\ref{fig:4} bottom left panel shows the results for this setup. The green and red dots represent the results for [Fe/H]=$-0.25$ and [Fe/H]= $+0.25$, respectively; the 3$\sigma$ limits are shown as the green and red dashed lines, and the blue crosses show our default results. We measure a cooling anomaly ranging from 0.94$^{+1.20}_{-1.06}$ Gyr to 1.19$^{+1.02}_{-0.97}$ Gyr for this metallicity range, [Fe/H]=$-0.25$ to $+0.25$. Hence, the choice of the progenitor star metallicity does not significantly impact our results.

Lastly, to test the effects of the electron conductive opacities on the age estimates, we compare our default results that use MESA models from \citet{bauer2023} and conductive opacities from \citet{blouin2020b} with a MESA model with the same properties but using the conductive opacities from \citet{cassisi2007}. This comparison is displayed as the green dots in the bottom right panel of Figure~\ref{fig:4}. The green dashed lines mark the 3$\sigma$ limits. We do not observe any significant differences in age estimates, as all our WDs have cooling ages less than 4 Gyr \citep[see Fig. 6 in ][]{bauer2023}. However, the crystallized core fractions change under the assumption of different conductive opacities, decreasing by about 10\% for the \citet{cassisi2007} opacities. This is represented as the dotted lines connecting blue crosses and green dots. Nevertheless, the cooling anomaly, $\Delta$Age = 1.10$^{+1.20}_{-1.08}$ Gyr, is similar to the one in our default model.

\subsection{$^{22}$Ne Distillation}
\label{sec:4.3}

\citet{blouin2021a} demonstrated that the exact shape of the C/O/Ne phase diagram is critical, as it determines the qualitative outcome of the distillation process and whether a $^{22}$Ne-rich core or shell is formed. The latter results in a significantly smaller amount of gravitational energy release and a smaller cooling delay. This cooling delay is expected to take place after $\sim$60\% of the core is crystallized, assuming an initially homogeneous C/O profile with X(O) = 0.60. For typical $^{22}$Ne mass fractions of 1.4\%, the predicted cooling delays are 1.6-2 Gyr for $M=0.6-1.0~M_{\odot}$ WDs. 

Stellar mergers may produce C/O WDs with significantly larger amounts of heavy neutron-rich impurities like $^{22}$Ne. \citet{bedard24} present the expected cooling delays from $^{22}$Ne distillation in ultramassive WDs that are enriched in $^{22}$Ne. They estimate that these stars will form a $^{22}$Ne-rich central core, and the resulting distillation process can lead to cooling delays of order 7-13 Gyr depending on the stellar mass (see their Figure~4a). They conclude that the cooling delay provided by $^{22}$Ne distillation can indeed explain the ultramassive WD cooling anomaly discovered by \citet[][]{cheng2019}. The range of masses analyzed in that work does not overlap with our sample.

There is only a single pair in our sample with a WD that is close to the $\sim60$\% crystallized core fraction. J1113+3238 is the most interesting, as one of its members has a C/O crystallized core fraction of 52\%. The parameters for each binary member of this system were discussed in Section \ref{sec:4.1}. Most importantly, we measure a relatively small cooling anomaly of $\Delta$Age = 0.82$^{+2.99}_{-0.86}$ Gyr for the crystallized WD in this system. Since the models used in our study include the extra energy from the latent heat of crystallization and C/O phase separation, we can isolate the effects of $^{22}$Ne distillation. Although the large age uncertainties in our study do not allow us to strongly constrain this effect, we can provide a firm empirical upper limit of 3.6 Gyr ($3\sigma$ limit) on the cooling delays associated with the distillation of $^{22}$Ne or other neutron-rich impurities for WDs with masses between 0.6-0.9 M$_{\odot}$.

\subsection{Convective Coupling}

\begin{figure}
	\includegraphics[width=0.47\textwidth]{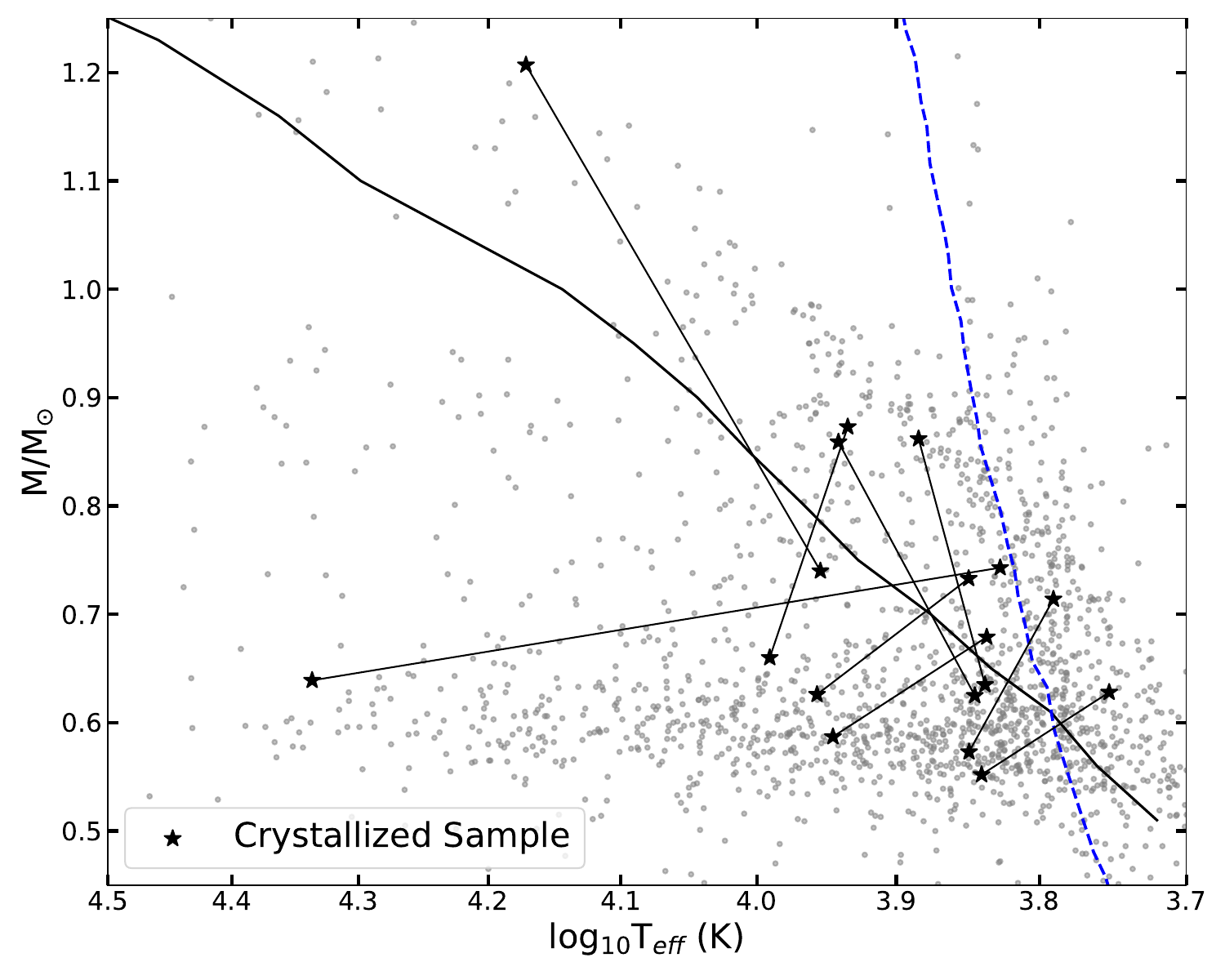}
    \caption{White Dwarfs masses as a function of the effective temperature for our sample of wide double white dwarfs with a crystallized member (black stars). The grey dots in the background are the 100 pc WD sample from \citet{kilic2020}, the black solid line represents the onset of crystallization based on \citet{bauer2023} C/O models (M$\leq$1.03) and \citet{camisassa2019} O/Ne models  (M$\geq$1.10), and the blue dashed line shows the onset of convective coupling based on \citet{bedard2020} models.}
   \label{fig:5}
\end{figure}

Another important process that affects the cooling rate in WDs is the onset of convective coupling. For massive WDs, convective coupling happens long after the start of crystallization. Hence, its impact on the evolutionary timescales is distinct from crystallization. On the other hand, for regular WDs around 0.6 M$_{\odot}$, both of these processes overlap, and therefore convective coupling can mask the cooling delays from crystallization alone \citep[e.g.,][]{fontaine2001}. Figure~\ref{fig:5} displays the effective temperatures and masses for our sample of DWDs with a crystallized member (black stars) along with the 100 pc WD sample from \citet[][grey dots]{kilic2020}. The solid black line represents the onset of crystallization for C/O core \citep[$M\leq1.03~M_{\odot}$,][]{bauer2023} and O/Ne core \citep[$M\geq1.10~M_{\odot}$][]{camisassa2019} WDs, whereas the blue dashed line shows the onset of convective coupling \citep{bedard2020}. Based on this figure, two of the WDs in our analysis have gone through the convective coupling process, J105242.54+283252 and J231939.16-035857. If we further exclude these two pairs from our sample of crystallized WDs, the resulting $\Delta$Age value is slightly lower ($\sim$0.3 Gyr). However, given the relatively large errors in our $\Delta$Age measurement, this change is insignificant.

\subsection{No Additional Cooling Delays in a Crystallized O/Ne Core White Dwarf}
\label{sec:4.4}

There is only one DWD in our sample with a member that is more than 60\% crystallized: J115937.82+134408.7. This is a DA WD with $T_{\rm eff}= 14833 \pm 490$ K and $M= 1.21 \pm 0.01~M_{\odot}$, and it has a core that is 88\% crystallized. Assuming that the two stars did not interact during their evolution, J115937.82+134408.7 likely has an O/Ne core, given its high mass. This is consistent with the fact that both stars in this binary appear to be normal DA WDs with disk kinematics. Therefore, we use O/Ne core models from \citet{camisassa2019} for the massive WD in this binary.

This type of WD crystallizes following a standard scenario where the solid crystals are heavier than the liquid, and therefore, they sink to the underlying solid core \citep[][]{camisassa2019, blouin2021c}. The amount of energy released in this case, both from latent heat and gravitational energy, is significantly smaller than the $^{22}$Ne distillation process \citep[see Fig.~2b in][]{bedard24}. Hence, we do not expect to find significant cooling anomalies in O/Ne core WDs.  

The companion in this case, J115937.81+134413.9, is a regular C/O core WD with mass $0.74~M_{\odot}$ that is not crystallized. We find a cooling anomaly of $0.18^{+0.22}_{-0.13}$ Gyr for this system. Hence, there is no evidence of any extra cooling delays in this likely O/Ne core WD with a core that is significantly crystallized. Since $^{22}$Ne distillation cannot take place in O/Ne cores \citep{camisassa2019,bedard24}, this result is not surprising, but it is re-affirming that the observations find no significant additional cooling delays in this system. This should be compared to the $\sim10$ Gyr cooling delays inferred for similarly massive WDs, including massive DQs and DAs, on the Q-branch \citep{cheng2019,bedard24,kilic24}.

\subsection{The Way Forward: Subgiant + Crystallized White Dwarf Binaries}
\label{sec:4.5}

The ultimate limitation in testing the cooling physics using DWDs is the precision of the age measurements for WDs. We showed here that even after restricting our sample to WDs with masses above 0.63 M$_{\odot}$ \citep{heintz2022}, we still deal with large age uncertainties since it is inherently difficult to obtain precise total ages for typical WDs with $M\sim0.6~M_{\odot}$. The ideal systems for characterizing the cooling delays introduced by $^{22}$Ne distillation would involve white dwarfs with well-determined ages and cores that are more than $\sim$60\% crystallized.

\citet{barrientos2021} presented an alternate sample where the ages of WDs can be determined using their subgiant companions. They used subgiant + WD binaries to constrain the initial-final mass relation by relying on the precise age dating of the subgiant companions. A similar approach can be considered using main sequence + WD binaries as done in \citet{rebassa2023}.

We selected a sample of $\sim$60 binaries from \citet{elbadry2021} where the primary star is a subgiant and the secondary is a crystallized WD based on the \citet{bedard2020} models. Table~\ref{tab:A2} presents a list of these newly identified pairs with \citet{gf2021} parameters for the WDs. This alternate sample has the potential to provide better system ages for these pairs containing crystallized WDs. We encourage and plan on obtaining follow-up observations of the most interesting systems in this sample to characterize their system parameters and empirically constrain the cooling delays from $^{22}$Ne distillation and other associated effects. 

\vskip 8mm
\section{Summary and Conclusions}
\label{sec:5}

We present a detailed model atmosphere analysis of 29 wide double white dwarfs, including 9 systems where one of the members is crystallized. We selected these high-confidence binaries from \citet{elbadry2021} and obtained spectroscopic follow-up for the targets with no spectral information in the literature. Our final sample includes white dwarfs with crystallized core fractions of 10\% to 88\%.

We used a Bayesian approach to calculate the cooling age, progenitor lifetime, and the total age of each star, and then we used these ages to search for cooling anomalies in the crystallized members. We take the total age of the non-crystallized star as the true age of each binary. We find a cooling anomaly of $\Delta$Age= 1.13$^{+1.20}_{-1.07}$ Gyr for the 8 binaries with crystallized white dwarfs and C/O cores. The same analysis on the control sample gives $\Delta$Age = $-0.03 \pm$ 0.15 Gyr, indicating that this method gives reliable results. 

Our results are consistent with the null hypothesis; given the relatively large errors, we do not find a significant cooling anomaly in our sample of 8 wide double white dwarf systems with crystallized C/O core members. More importantly, we can rule out cooling anomalies larger than 3.6 Gyr due to $^{22}$Ne distillation and other neutron-rich impurities at the 99.7\% ($3\sigma$) confidence level for white dwarfs of masses between 0.6-0.9 M$_{\odot}$.

The use of wide binaries with at least one white dwarf component provides a new way to test the cooling physics of these objects. In particular, we present a sample of $\sim$60 binaries composed of a subgiant star with a crystallized white dwarf companion. The subgiant can be age-dated precisely to estimate the total system age, which can then be used to infer any cooling anomalies present in the crystallized white dwarf companion. Future analysis of these pairs can help us obtain and improve age measurements and, therefore, bring us closer to empirically constraining the cooling delays in white dwarfs.

\section*{Acknowledgements}

M.B. acknowledges the invaluable discussions, feedback, and suggestions by collaborators and referees. This work is supported in part by the NSF under grant  AST-2205736, the NASA under grants 80NSSC22K0479, 80NSSC24K0380, and 80NSSC24K0436, the NSERC Canada, the Fund FRQ-NT (Qu\'ebec), the Canadian Institute for Theoretical Astrophysics (CITA) National Fellowship Program, and by the Smithsonian Institution.

Based on observations obtained at the MMT Observatory, a joint facility of the Smithsonian Institution and the University of Arizona.

Based on observations obtained at the international Gemini Observatory, a program of NSF's NOIRLab, which is managed by the Association of Universities for Research in Astronomy (AURA) under a cooperative agreement with the National Science Foundation on behalf of the Gemini Observatory partnership: the National Science Foundation (United States), National Research Council (Canada), Agencia Nacional de Investigaci\'{o}n y Desarrollo (Chile), Ministerio de Ciencia, Tecnolog\'{i}a e Innovaci\'{o}n (Argentina), Minist\'{e}rio da Ci\^{e}ncia, Tecnologia, Inova\c{c}\~{o}es e Comunica\c{c}\~{o}es (Brazil), and Korea Astronomy and Space Science Institute (Republic of Korea).

\vskip 20mm
\appendix

\section{Data Tables and Figures}
\counterwithin{table}{section}
\counterwithin{figure}{section}
\setcounter{table}{0}
\setcounter{figure}{0}
\label{sec:appendixA}

\begin{figure*}[h!]
\centering
        \includegraphics[width=\textwidth]{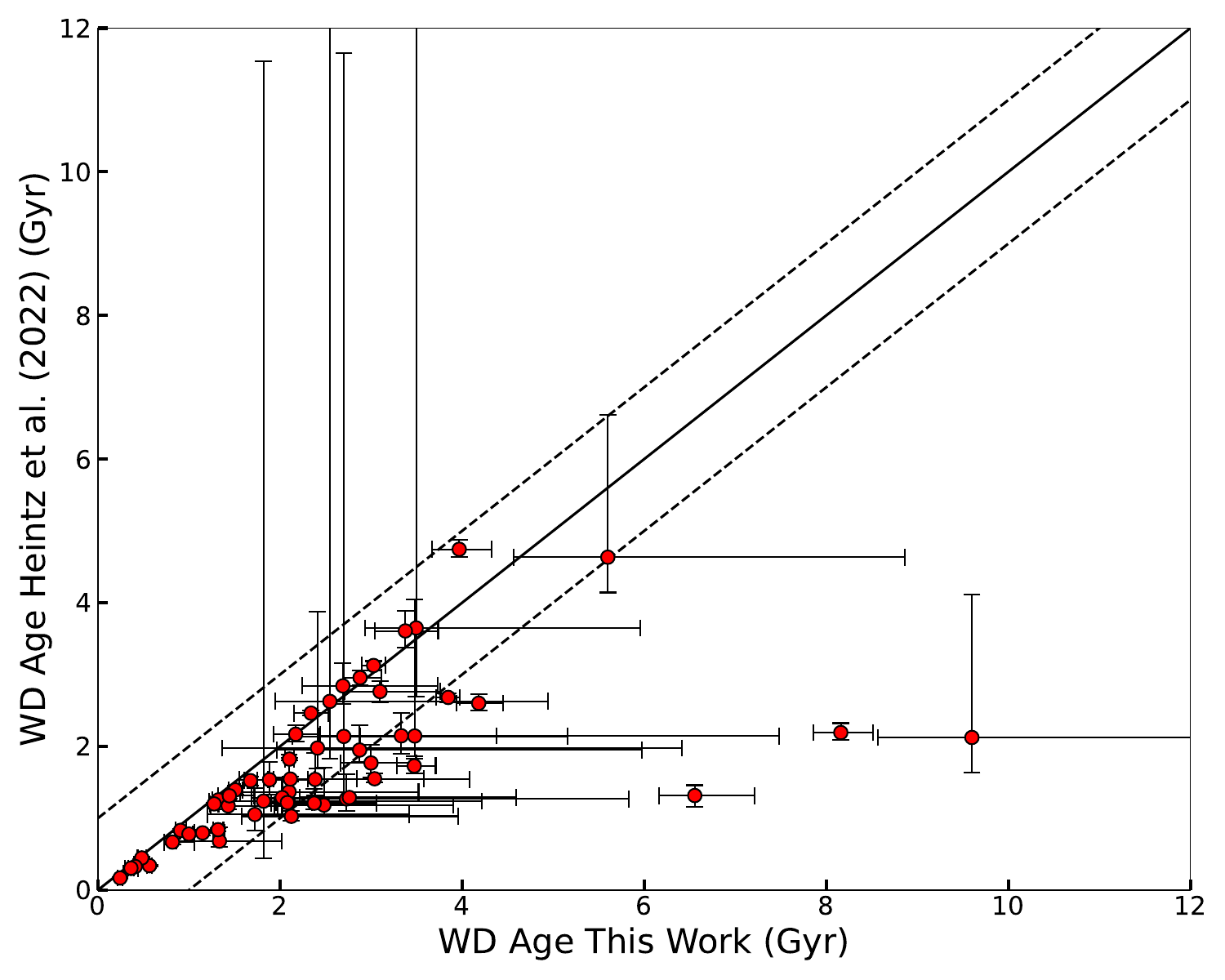}
    \caption{Total age comparison between this work and \citet{heintz2022}. The largest differences are seen for J105242.57+283255.3 (DQ), J123647.42+682502.9 (DQ), and J231938.65-035833.1 (DC).}
    \label{fig:A1}
\end{figure*}

In this section, we show additional tables and a figure as part of this work. In Table~\ref{tab:A1}, we provide the physical parameters of the control sample used in this work. Table~\ref{tab:A2} and Figure~\ref{fig:A1} present an age comparison of the binary systems analyzed here and in \citet{heintz2022}. In Table~\ref{tab:A3}, we provide a list of candidate Subgiant + crystallized WD binaries for future studies. 

\begin{deluxetable*}{crrrrrrrcc}[htp]
\tabletypesize{\footnotesize}
\tablecolumns{10} \tablewidth{0pt}
\tablecaption{The physical parameters for the Double White Dwarfs in our Control Sample.} 
\label{tab:A1}
\tablehead{\colhead{Object name} & \colhead{{\emph Gaia} Source ID} & \colhead{Type} & \colhead{T$_{\text{eff}}$} & \colhead{log(g)} & \colhead{M$_{\text{WD}}$ } & \colhead{$\tau_{\text{cooling}}$ } & \colhead{$\tau_{\text{progenitor}}$} & \colhead{Total Age} & \colhead{$\Delta$ Age$^a$} \\
& \colhead{(DR3)} & & \colhead{(K)}  & \colhead{(cm s$^{-1}$)} & \colhead{(M$_{\odot}$)} &  \colhead{(Gyr)} & \colhead{(Gyr)} & \colhead{(Gyr)} & \colhead{(Gyr)} }
\startdata
\hline\hline
J003051.80+181046.0 & 2794800060629297152 & DA & 14366 $\pm$ 292 & 8.28 $\pm$ 0.05 & 0.75 $\pm$ 0.03 & 0.34$^{+0.04}_{-0.04}$ & 0.47$^{+0.26}_{-0.11}$ & 0.82$^{+0.24}_{-0.09}$ & \multirow{2}{*}{0.40$^{+0.58}_{-0.32}$}\\
J003051.75+181053.7 & 2794800056333855232 & DA & 13633 $\pm$ 234 & 8.20 $\pm$ 0.05 & 0.70 $\pm$ 0.03 & 0.33$^{+0.04}_{-0.03}$ & 0.82$^{+0.57}_{-0.30}$ & 1.15$^{+0.56}_{-0.27}$ & \\
\hline
J080333.78-090705.8 & 3039398650000089472 & DA & 11247 $\pm$ 130 & 8.07 $\pm$ 0.02 & 0.63 $\pm$ 0.01 & 0.46$^{+0.02}_{-0.02}$ & 2.25$^{+3.14}_{-0.89}$ & 2.73$^{+3.10}_{-0.91}$ & \multirow{2}{*}{0.92$^{+3.51}_{-1.57}$}\\
J080333.92-090658.4 & 3039398650000089600 & DC & 7933 $\pm$ 177 & 8.11 $\pm$ 0.03 & 0.65 $\pm$ 0.02 & 1.19$^{+0.10}_{-0.09}$ & 1.68$^{+2.35}_{-0.63}$ & 2.87$^{+2.28}_{-0.61}$ & \\
\hline
J082730.72-021618.5 & 3072961074934467200 & DA & 27037 $\pm$ 980 & 8.60 $\pm$ 0.04 & 0.97 $\pm$ 0.02 & 0.10$^{+0.02}_{-0.02}$ & 0.14$^{+0.03}_{-0.02}$ & 0.24$^{+0.03}_{-0.03}$ & \multirow{2}{*}{0.17$^{+0.08}_{-0.07}$}\\
J082730.58-021620.1 & 3072961070640767488 & DA & 24501 $\pm$ 930 & 8.33 $\pm$ 0.04 & 0.81 $\pm$ 0.03 & 0.07$^{+0.02}_{-0.01}$ & 0.34$^{+0.08}_{-0.06}$ & 0.41$^{+0.07}_{-0.06}$ & \\
\hline
J085917.36+425031.6 & 1008929564913828224 & DA & 9707 $\pm$ 153 & 8.14 $\pm$ 0.04 & 0.66 $\pm$ 0.03 & 0.73$^{+0.06}_{-0.06}$ & 1.36$^{+1.47}_{-0.58}$ & 2.10$^{+1.44}_{-0.54}$ & \multirow{2}{*}{0.88$^{+3.26}_{-1.27}$}\\
J085917.23+425027.4 & 1008929569208837376 & DA & 11106 $\pm$ 117 & 8.09 $\pm$ 0.04 & 0.64 $\pm$ 0.02 & 0.48$^{+0.03}_{-0.03}$ & 2.00$^{+3.14}_{-0.83}$ & 2.49$^{+3.15}_{-0.81}$ & \\
\hline
J092513.18+160145.4 & 630770819920096768 & DA & 21766 $\pm$ 442 & 8.83 $\pm$ 0.01 & 1.10 $\pm$ 0.01 & 0.49$^{+0.03}_{-0.03}$ & 0.08$^{+0.01}_{-0.01}$ & 0.57$^{+0.03}_{-0.03}$ & \multirow{2}{*}{-0.20$^{+0.05}_{-0.05}$}\\
J092513.48+160144.1 & 630770819920096640 & DA & 24050 $\pm$ 482 & 8.40 $\pm$ 0.02 & 0.85 $\pm$ 0.01 & 0.09$^{+0.01}_{-0.01}$ & 0.27$^{+0.05}_{-0.04}$ & 0.36$^{+0.05}_{-0.04}$ & \\
\hline
J100244.88+360629.6 & 795886439568266368 & DA & 10428 $\pm$ 176 & 8.23 $\pm$ 0.06 & 0.71 $\pm$ 0.03 & 0.70$^{+0.09}_{-0.07}$ & 0.72$^{+0.61}_{-0.25}$ & 1.43$^{+0.55}_{-0.20}$ & \multirow{2}{*}{1.78$^{+4.09}_{-1.14}$}\\
J100245.86+360653.3 & 795886439568268032 & DA & 9471 $\pm$ 135 & 8.09 $\pm$ 0.06 & 0.63 $\pm$ 0.03 & 0.72$^{+0.07}_{-0.06}$ & 2.33$^{+4.11}_{-1.12}$ & 3.04$^{+4.09}_{-1.07}$ & \\
\hline
J111020.98+451801.7 & 782193985044906752 & DA & 20511 $\pm$ 348 & 8.11 $\pm$ 0.02 & 0.67 $\pm$ 0.01 & 0.08$^{+0.01}_{-0.01}$ & 1.25$^{+0.69}_{-0.47}$ & 1.33$^{+0.69}_{-0.47}$ & \multirow{2}{*}{1.21$^{+2.66}_{-0.95}$}\\
J111016.68+451736.3 & 782194019404645632 & DA & 12992 $\pm$ 180 & 8.07 $\pm$ 0.01 & 0.64 $\pm$ 0.01 & 0.32$^{+0.01}_{-0.01}$ & 2.07$^{+2.62}_{-0.75}$ & 2.38$^{+2.65}_{-0.74}$ & \\
\hline
J122717.81+563821.7 & 1574653689250595072 & DA & 8543 $\pm$ 225 & 8.13 $\pm$ 0.05 & 0.65 $\pm$ 0.03 & 0.99$^{+0.11}_{-0.09}$ & 1.54$^{+2.46}_{-0.66}$ & 2.55$^{+2.36}_{-0.60}$ & \multirow{2}{*}{-0.04$^{+2.91}_{-1.31}$}\\
J122717.42+563825.6 & 1574653689250360576 & DA & 17013 $\pm$ 422 & 8.08 $\pm$ 0.03 & 0.65 $\pm$ 0.02 & 0.15$^{+0.01}_{-0.01}$ & 1.66$^{+2.21}_{-0.61}$ & 1.81$^{+2.20}_{-0.61}$ & \\
\hline
J131332.14+203039.6 & 3940068410255312768 & DA & 12670 $\pm$ 154 & 8.35 $\pm$ 0.01 & 0.81 $\pm$ 0.01 & 0.56$^{+0.03}_{-0.02}$ & 0.35$^{+0.06}_{-0.05}$ & 0.91$^{+0.06}_{-0.05}$ & \multirow{2}{*}{0.41$^{+0.46}_{-0.31}$}\\
J131332.56+203039.4 & 3940068414551140608 & DA & 12789 $\pm$ 137 & 8.16 $\pm$ 0.01 & 0.69 $\pm$ 0.01 & 0.38$^{+0.01}_{-0.01}$ & 0.94$^{+0.45}_{-0.31}$ & 1.32$^{+0.46}_{-0.30}$ & \\
\hline
J170355.91+330438.4 & 1337576648471644800 & DA & 9932 $\pm$ 126 & 8.24 $\pm$ 0.03 & 0.73 $\pm$ 0.01 & 0.85$^{+0.05}_{-0.05}$ & 0.59$^{+0.29}_{-0.14}$ & 1.45$^{+0.27}_{-0.13}$ & \multirow{2}{*}{0.71$^{+1.79}_{-0.57}$}\\
J170356.77+330435.7 & 1337576648471644288 & DA & 11120 $\pm$ 131 & 8.10 $\pm$ 0.02 & 0.65 $\pm$ 0.01 & 0.50$^{+0.02}_{-0.02}$ & 1.59$^{+1.81}_{-0.54}$ & 2.08$^{+1.78}_{-0.53}$ & \\
\hline
J211507.42-074134.5 & 6898489884295412352 & DA & 10622 $\pm$ 51 & 8.42 $\pm$ 0.01 & 0.84 $\pm$ 0.01 & 0.99$^{+0.02}_{-0.02}$ & 0.28$^{+0.05}_{-0.04}$ & 1.28$^{+0.05}_{-0.05}$ & \multirow{2}{*}{0.83$^{+0.47}_{-0.25}$}\\
J211507.39-074151.5 & 6898489884295407488 & DA & 8131 $\pm$ 36 & 8.19 $\pm$ 0.01 & 0.70 $\pm$ 0.01 & 1.27$^{+0.03}_{-0.02}$ & 0.83$^{+0.46}_{-0.24}$ & 2.10$^{+0.47}_{-0.24}$ & \\
\hline
J222236.56-082806.0 & 2616210918121365760 & DA & 11389 $\pm$ 146 & 8.10 $\pm$ 0.01 & 0.65 $\pm$ 0.01 & 0.47$^{+0.02}_{-0.02}$ & 1.66$^{+1.87}_{-0.55}$ & 2.12$^{+1.84}_{-0.55}$ & \multirow{2}{*}{1.24$^{+3.57}_{-1.62}$}\\
J222236.30-082807.9 & 2616210922414728960 & DA & 15704 $\pm$ 165 & 8.04 $\pm$ 0.01 & 0.63 $\pm$ 0.01 & 0.18$^{+0.01}_{-0.01}$ & 2.55$^{+3.36}_{-1.14}$ & 2.74$^{+3.35}_{-1.15}$ & \\
\hline
J222301.62+220131.3 & 1874954641491354624 & DA & 19185 $\pm$ 245 & 8.40 $\pm$ 0.01 & 0.84 $\pm$ 0.01 & 0.20$^{+0.01}_{-0.01}$ & 0.29$^{+0.05}_{-0.05}$ & 0.48$^{+0.05}_{-0.05}$ & \multirow{2}{*}{0.52$^{+0.28}_{-0.15}$}\\
J222301.72+220124.9 & 1874954645786146304 & DA & 13401 $\pm$ 91 & 8.22 $\pm$ 0.01 & 0.72 $\pm$ 0.00 & 0.38$^{+0.01}_{-0.01}$ & 0.62$^{+0.28}_{-0.15}$ & 1.00$^{+0.28}_{-0.15}$ & \\
\hline
J231941.34+342614.2 & 1911420636118031744 & DA & 16601 $\pm$ 351 & 8.08 $\pm$ 0.02 & 0.65 $\pm$ 0.01 & 0.16$^{+0.01}_{-0.01}$ & 1.55$^{+1.72}_{-0.52}$ & 1.72$^{+1.69}_{-0.52}$ & \multirow{2}{*}{1.24$^{+2.98}_{-1.29}$}\\
J231941.44+342609.3 & 1911420636119326976 & DA & 13882 $\pm$ 217 & 8.06 $\pm$ 0.02 & 0.64 $\pm$ 0.01 & 0.27$^{+0.01}_{-0.01}$ & 2.10$^{+2.77}_{-0.76}$ & 2.38$^{+2.75}_{-0.77}$ & \\
\hline\hline
\enddata
\end{deluxetable*}

\begin{deluxetable*}{ccccc}[htp]
\tabletypesize{\footnotesize}
\tablecolumns{5} \tablewidth{0pt}
\tablecaption{Age Comparison: This Work vs \citet{heintz2022}} 
\label{tab:A2}
\tablehead{\colhead{Object name} & \colhead{{\emph Gaia} Source ID} & \colhead{Type} & \colhead{Total Age This Work } & \colhead{ Total Age \citet{heintz2022}} \\
& \colhead{(DR3)} & & \colhead{(Gyr)} & \colhead{(Gyr)} }
\startdata
\hline\hline
J002925.29+001559.7 & 2543654008364462848 & DC & 2.34$^{+0.19}_{-0.18}$ & 2.46$^{+0.04}_{-0.03}$\\
J002925.62+001552.7 & 2543653978300417024 & DA & 2.11$^{+1.59}_{-0.54}$ & 1.55$^{+0.56}_{-0.25}$\\
\hline
J081427.92+013325.6 & 3089916403229120640 & DA & 1.89$^{+0.49}_{-0.33}$ & 1.54$^{+0.25}_{-0.20}$\\
J081427.83+013318.7 & 3089916398933726592 & DA & 3.00$^{+5.03}_{-0.95}$ & 1.77$^{+31.10}_{-0.34}$\\
\hline
J102142.06+394225.4 & 804040486519166976 & DA & 2.69$^{+1.04}_{-0.45}$ & 2.84$^{+0.32}_{-0.25}$\\
J102141.29+394215.5 & 804040108562044288 & DA & 3.33$^{+5.00}_{-1.05}$ & 2.15$^{+2.12}_{-0.23}$\\
\hline
J105242.54+283252.0 & 731411283875749760 & DA & 3.97$^{+0.35}_{-0.30}$ & 4.75$^{+0.13}_{-0.11}$\\
J105242.57+283255.3 & 731411283874346240 & DQ & 8.16$^{+4.34}_{-3.54}$ & 2.19$^{+0.13}_{-0.17}$\\
\hline
J111322.48+323859.0 & 757911988004305280 & DA & 3.03$^{+0.13}_{-0.13}$ & 3.13$^{+0.06}_{-0.07}$\\
J111319.38+323818.0 & 757911884925087104 & DA & 3.85$^{+3.03}_{-0.85}$ & 2.68$^{+0.45}_{-0.22}$\\
\hline
J115749.13+313931.0 & 4026615235380699392 & DAH & 2.17$^{+0.27}_{-0.25}$ & 2.17$^{+0.12}_{-0.11}$\\
J115749.39+313931.0 & 4026615235380699520 & DA & 4.18$^{+4.54}_{-1.26}$ & 2.60$^{+1.17}_{-0.29}$\\
\hline
J115937.82+134408.7 & 3920275276810355200 & DA & 1.51$^{+0.07}_{-0.07}$ & 1.39$^{+0.05}_{-0.07}$\\
J115937.81+134413.9 & 3920275276810355072 & DA & 1.68$^{+0.21}_{-0.10}$ & 1.52$^{+0.04}_{-0.04}$\\
\hline
J123647.95+682501.6 & 1682366418152856448 & DA & 3.10$^{+0.66}_{-0.39}$ & 2.76$^{+0.14}_{-0.15}$\\
J123647.42+682502.9 & 1682554091043762560 & DQ & 6.56$^{+4.68}_{-3.10}$ & 1.32$^{+0.03}_{-0.06}$\\
\hline
J134739.13+251228.9 & 1444442547261998464 & DA & 3.50$^{+2.46}_{-0.56}$ & 3.65$^{+9.51}_{-0.96}$\\
J134737.58+251233.1 & 1444442512902260864 & DA & 2.70$^{+2.63}_{-0.63}$ & 2.14$^{+4.52}_{-0.41}$\\
\hline
J135834.36+263345.2 & 1450779342012324224 & DA & 3.38$^{+0.36}_{-0.34}$ & 3.61$^{+0.28}_{-0.23}$\\
J135834.66+263343.1 & 1450779346306149760 & DA & 2.02$^{+2.64}_{-0.72}$ & 1.29$^{+1.85}_{-0.45}$\\
\hline
J185722.65+781332.1 & 2293210651405001216 & DA & 3.48$^{+4.00}_{-1.05}$ & 2.15$^{+1.90}_{-0.29}$\\
J185722.00+781332.2 & 2293210651402924160 & DA & 2.41$^{+1.79}_{-0.39}$ & 1.98$^{+0.19}_{-0.11}$\\
\hline
J213102.82+083203.8 & 1741031891851063424 & DA & 2.88$^{+0.23}_{-0.19}$ & 2.96$^{+0.10}_{-0.10}$\\
J213103.09+083202.6 & 1741031896140411648 & DA & 3.48$^{+3.51}_{-1.24}$ & 1.73$^{+0.48}_{-0.23}$\\
\hline
J225932.73+140444.2 & 2815944352131513472 & DAH & 1.32$^{+0.08}_{-0.07}$ & 1.26$^{+0.04}_{-0.03}$\\
J225932.21+140439.2 & 2815944352131513088 & DA & 0.36$^{+0.06}_{-0.05}$ & 0.30$^{+0.02}_{-0.02}$\\
\hline
J231939.16-035857.8 & 2634002940402436480 & DAH & 5.60$^{+3.26}_{-1.03}$ & 4.63$^{+1.99}_{-0.49}$\\
J231938.65-035833.1 & 2634003146560869248 & DC & 9.60$^{+3.57}_{-3.67}$ & 2.12$^{+0.06}_{-0.08}$\\
\hline\hline
\enddata
\end{deluxetable*}

\begin{deluxetable*}{ccccc}[htp]
\tabletypesize{\footnotesize}
\tablecolumns{5} \tablewidth{10pt} 
\tablecaption{Continuation Age Comparison: This Work vs \citet{heintz2022}} 
\addtocounter{table}{-1}
\tablehead{\colhead{Object name} & \colhead{{\emph Gaia} Source ID} & \colhead{Type} & \colhead{Total Age This Work } & \colhead{ Total Age \citet{heintz2022}} \\
& \colhead{(DR3)} & & \colhead{(Gyr)} & \colhead{(Gyr)} }
\startdata
\hline\hline
J003051.80+181046.0 & 2794800060629297152 & DA & 0.82$^{+0.24}_{-0.09}$ & 0.67$^{+0.05}_{-0.04}$\\
J003051.75+181053.7 & 2794800056333855232 & DA & 1.15$^{+0.56}_{-0.28}$ & 0.80$^{+0.10}_{-0.06}$\\
\hline
J080333.78-090705.8 & 3039398650000089472 & DA & 2.73$^{+3.10}_{-0.91}$ & 1.27$^{+0.34}_{-0.17}$\\
J080333.92-090658.4 & 3039398650000089600 & DC & 2.88$^{+2.29}_{-0.61}$ & 1.95$^{+0.06}_{-0.07}$\\
\hline
J082730.72-021618.5 & 3072961074934467200 & DA & 0.24$^{+0.03}_{-0.03}$ & 0.17$^{+0.02}_{-0.01}$\\
J082730.58-021620.1 & 3072961070640767488 & DA & 0.41$^{+0.07}_{-0.06}$ & 0.33$^{+0.03}_{-0.04}$\\
\hline
J085917.36+425031.6 & 1008929564913828224 & DA & 2.10$^{+1.42}_{-0.54}$ & 1.37$^{+0.52}_{-0.13}$\\
J085917.23+425027.4 & 1008929569208837376 & DA & 2.48$^{+3.15}_{-0.81}$ & 1.18$^{+0.60}_{-0.17}$\\
\hline
J092513.18+160145.4 & 630770819920096768 & DA & 0.57$^{+0.03}_{-0.03}$ & 0.34$^{+0.01}_{-0.01}$\\
J092513.48+160144.1 & 630770819920096640 & DA & 0.36$^{+0.05}_{-0.04}$ & 0.31$^{+0.02}_{-0.03}$\\
\hline
J100244.88+360629.6 & 795886439568266368 & DA & 1.43$^{+0.55}_{-0.20}$ & 1.17$^{+0.07}_{-0.05}$\\
J100245.86+360653.3 & 795886439568268032 & DA & 3.04$^{+4.09}_{-1.07}$ & 1.55$^{+1.44}_{-0.23}$\\
\hline
J111020.98+451801.7 & 782193985044906752 & DA & 1.33$^{+0.69}_{-0.47}$ & 0.68$^{+0.20}_{-0.08}$\\
J111016.68+451736.3 & 782194019404645632 & DA & 2.38$^{+2.64}_{-0.73}$ & 1.21$^{+0.36}_{-0.20}$\\
\hline
J122717.81+563821.7 & 1574653689250595072 & DA & 2.55$^{+2.40}_{-0.60}$ & 2.63$^{+10.31}_{-0.79}$\\
J122717.42+563825.6 & 1574653689250360576 & DA & 1.82$^{+2.20}_{-0.61}$ & 1.24$^{+2.26}_{-0.40}$\\
\hline
J131332.14+203039.6 & 3940068410255312768 & DA & 0.91$^{+0.06}_{-0.05}$ & 0.83$^{+0.02}_{-0.02}$\\
J131332.56+203039.4 & 3940068414551140608 & DA & 1.32$^{+0.46}_{-0.30}$ & 0.84$^{+0.04}_{-0.04}$\\
\hline
J170355.91+330438.4 & 1337576648471644800 & DA & 1.44$^{+0.27}_{-0.13}$ & 1.31$^{+0.05}_{-0.04}$\\
J170356.77+330435.7 & 1337576648471644288 & DA & 2.08$^{+1.78}_{-0.53}$ & 1.22$^{+0.27}_{-0.12}$\\
\hline
J211507.42-074134.5 & 6898489884295412352 & DA & 1.28$^{+0.05}_{-0.05}$ & 1.20$^{+0.02}_{-0.02}$\\
J211507.39-074151.5 & 6898489884295407488 & DA & 2.10$^{+0.46}_{-0.24}$ & 1.82$^{+0.06}_{-0.04}$\\
\hline
J222236.56-082806.0 & 2616210918121365760 & DA & 2.12$^{+1.83}_{-0.55}$ & 1.03$^{+0.08}_{-0.06}$\\
J222236.30-082807.9 & 2616210922414728960 & DA & 2.76$^{+3.33}_{-1.17}$ & 1.29$^{+0.31}_{-0.27}$\\
\hline
J222301.62+220131.3 & 1874954641491354624 & DA & 0.48$^{+0.05}_{-0.05}$ & 0.45$^{+0.01}_{-0.01}$\\
J222301.72+220124.9 & 1874954645786146304 & DA & 1.00$^{+0.28}_{-0.14}$ & 0.78$^{+0.02}_{-0.02}$\\
\hline
J231941.34+342614.2 & 1911420636118031744 & DA & 1.72$^{+1.70}_{-0.52}$ & 1.05$^{+0.36}_{-0.22}$\\
J231941.44+342609.3 & 1911420636119326976 & DA & 2.39$^{+2.76}_{-0.78}$ & 1.54$^{+1.13}_{-0.36}$\\
\hline\hline
\enddata
\end{deluxetable*}
\begin{table*}
\centering
\def\arraystretch{0.75}
\setlength{\tabcolsep}{11.0pt}
\caption{Subgiant-White Dwarf Binaries with \citet{gf2021} parameters}
\label{tab:A3}
\begin{tabular}{c c c c c c}
\hline\hline
SG {\emph Gaia} Source ID & WD {\emph Gaia} Source ID & SG Gmag & WD Gmag & WD T$_{\text{eff}}$ & WD Mass  \\ 
(DR3) & (DR3) & (mag) & (mag) & (K) & (M$_{\odot}$)\\
\hline\hline
4300369510174214784 & 4300369510169386112 & 8.81 & 19.66 & 12912$\pm$3151 & 0.88$\pm$0.25\\
2336787015926880000 & 2336787423948155008 & 9.74 & 19.82 & 10624$\pm$1785 & 1.05$\pm$0.22\\
2347362256201579904 & 2347362359280261248 & 9.31 & 20.59 & 5393$\pm$714 & 0.74$\pm$0.37\\
4392051256453106816 & 4392051256451595520 & 7.93 & 19.56 & 9787$\pm$1344 & 0.93$\pm$0.22\\
6133033601555979648 & 6133033635916500608 & 5.61 & 16.18 & 6799$\pm$72 & 0.76$\pm$0.02\\
2119845400308940160 & 2119845361652709376 & 10.22 & 19.74 & 9260$\pm$1733 & 0.88$\pm$0.29\\
3830984079252585600 & 3830990156631488128 & 7.93 & 17.09 & 10431$\pm$232 & 0.88$\pm$0.03\\
3831183812411727616 & 3831183705037343744 & 7.72 & 18.95 & 8640$\pm$1671 & 0.87$\pm$0.29\\
5383783218258538112 & 5383783355697619840 & 11.03 & 20.15 & 10578$\pm$2019 & 0.93$\pm$0.3\\
6598562289267096448 & 6598562289266484480 & 8.35 & 20.43 & 5148$\pm$1119 & 0.82$\pm$0.77\\
2376772955294340992 & 2376773024012888832 & 9.81 & 20.07 & 10685$\pm$2125 & 1.19$\pm$0.19\\
6716346883984454400 & 6716346888277566080 & 10.14 & 19.75 & 11886$\pm$3455 & 0.99$\pm$0.29\\
2285623681871320320 & 2285623677575389824 & 10.01 & 20.37 & 11346$\pm$3863 & 0.95$\pm$0.43\\
4331764106188333184 & 4331764106187008000 & 7.45 & 19.10 & 9549$\pm$1056 & 0.92$\pm$0.17\\
1360094169868916224 & 1360094169867063168 & 9.22 & 20.54 & 5160$\pm$977 & 0.72$\pm$0.45\\
2033936051408038144 & 2033937086539674752 & 9.92 & 20.67 & 7640$\pm$1658 & 1.1$\pm$0.29\\
6212459675847218816 & 6212459710206960256 & 8.66 & 18.31 & 10230$\pm$568 & 0.83$\pm$0.08\\
2879667068210826752 & 2879667033851088896 & 6.40 & 17.77 & 6386$\pm$119 & 0.78$\pm$0.04\\
2090626428561466112 & 2090626428557671424 & 10.04 & 19.93 & 9118$\pm$2034 & 0.86$\pm$0.35\\
2027675161648624768 & 2027675092907621760 & 9.34 & 19.36 & 8146$\pm$948 & 0.83$\pm$0.2\\
554338410851013248 & 554338410851400576 & 10.91 & 20.52 & 10763$\pm$3263 & 1.07$\pm$0.47\\
1103672557732637568 & 1103673313648946176 & 8.58 & 18.97 & 11932$\pm$2263 & 1.08$\pm$0.14\\
1959369925890722688 & 1959369925890724352 & 8.88 & 19.51 & 9603$\pm$964 & 0.89$\pm$0.17\\
5768161698670783104 & 5768161728731408512 & 8.54 & 18.82 & 7483$\pm$443 & 0.89$\pm$0.09\\
2777818000458834816 & 2777818069178793216 & 9.05 & 20.11 & 5760$\pm$674 & 0.76$\pm$0.28\\
522824414744038656 & 522824517819155840 & 8.72 & 18.48 & 9697$\pm$938 & 0.85$\pm$0.14\\
4908881095134079744 & 4908884011415732608 & 9.13 & 19.33 & 9796$\pm$940 & 0.83$\pm$0.16\\
2245852830169628288 & 2245852834465778048 & 9.43 & 19.47 & 11546$\pm$3274 & 0.93$\pm$0.29\\
5531471338089757824 & 5531471342386268544 & 9.73 & 20.04 & 9566$\pm$2096 & 0.93$\pm$0.33\\
2550020872178776448 & 2550020872178380416 & 8.57 & 19.59 & 5755$\pm$682 & 0.74$\pm$0.26\\
6860685689529683072 & 6860685689529693312 & 7.41 & 18.73 & 9142$\pm$1267 & 0.91$\pm$0.2\\
5007222548993285376 & 5007222342833916032 & 9.81 & 19.81 & 7372$\pm$1112 & 0.8$\pm$0.27\\
5046646500480419328 & 5046645023010375040 & 9.95 & 19.87 & 8318$\pm$1438 & 0.86$\pm$0.28\\
2153552647245580544 & 2153552814748001792 & 9.29 & 19.82 & 10247$\pm$2200 & 1.11$\pm$0.2\\
816035200601793664 & 816035402464238592 & 9.45 & 20.33 & 7496$\pm$1330 & 0.82$\pm$0.35\\
4454777398385627904 & 4454777398385626368 & 9.30 & 19.03 & 7912$\pm$465 & 0.82$\pm$0.11\\
1485875894904733312 & 1485852491629542144 & 8.79 & 18.90 & 6641$\pm$250 & 0.74$\pm$0.07\\
3158711368312044800 & 3158709955270315776 & 8.94 & 19.46 & 9766$\pm$1785 & 1.02$\pm$0.24\\
4940783321935202944 & 4940783321935203200 & 9.14 & 18.83 & 18169$\pm$4619 & 1.22$\pm$0.09\\
508792687856955136 & 508792687857693952 & 9.92 & 20.34 & 9360$\pm$1488 & 1.0$\pm$0.25\\
1583002147896863360 & 1583002113536258304 & 8.73 & 20.48 & 5362$\pm$687 & 0.84$\pm$0.32\\
1591057651117375104 & 1591057681181908736 & 9.06 & 18.73 & 8010$\pm$284 & 0.87$\pm$0.06\\
5643049095188961024 & 5643047613413300736 & 10.59 & 19.84 & 11115$\pm$1586 & 0.86$\pm$0.23\\
4742627996346980352 & 4742627996345892096 & 8.93 & 20.28 & 6065$\pm$1232 & 1.05$\pm$0.3\\
4972150528953711872 & 4972150533247866240 & 9.95 & 20.46 & 7144$\pm$2109 & 1.07$\pm$0.88\\
77232655968062080 & 77232862125996160 & 9.41 & 19.72 & 7457$\pm$1380 & 0.97$\pm$0.27\\
4573348766683678208 & 4573348693667903616 & 8.69 & 19.88 & 6400$\pm$519 & 0.9$\pm$0.15\\
2231263620836219520 & 2231263616542725120 & 9.80 & 19.68 & 9333$\pm$1109 & 0.91$\pm$0.19\\
1044069612939035904 & 1044069612939334272 & 8.72 & 19.42 & 5881$\pm$289 & 0.75$\pm$0.11\\
4358360983226863616 & 4358361356891222784 & 8.85 & 20.10 & 4961$\pm$574 & 0.74$\pm$0.29\\
5606450613704556288 & 5606450579335008128 & 8.98 & 19.38 & 6935$\pm$400 & 0.76$\pm$0.12\\
2185083715854536320 & 2185083715855943552 & 11.16 & 20.42 & 9306$\pm$2503 & 0.91$\pm$0.45\\
4564212134294495360 & 4564212061278453248 & 8.35 & 18.86 & 7703$\pm$453 & 0.88$\pm$0.1\\
531850305690610688 & 531850374406289664 & 8.56 & 19.01 & 6337$\pm$473 & 0.75$\pm$0.14\\
4476374207640229888 & 4476374108856134656 & 11.24 & 19.71 & 12003$\pm$2607 & 0.86$\pm$0.29\\
5371977010298123264 & 5371977010295931392 & 9.96 & 19.81 & 8742$\pm$1447 & 0.79$\pm$0.29\\
1656598027422451456 & 1656598027420856448 & 9.19 & 20.01 & 6605$\pm$964 & 0.84$\pm$0.26\\
4280947595034458496 & 4280947973002608128 & 8.98 & 20.10 & 7635$\pm$1587 & 0.79$\pm$0.4\\
2939426178022960128 & 2939426173722815872 & 9.57 & 19.79 & 7256$\pm$893 & 0.85$\pm$0.23\\
4213545852929485056 & 4213545848628209152 & 9.91 & 19.56 & 8351$\pm$819 & 0.79$\pm$0.19\\
3177760888297880192 & 3177761094456310656 & 7.05 & 17.37 & 8403$\pm$502 & 0.92$\pm$0.08\\
2455561896956826112 & 2455561896955810432 & 8.79 & 19.56 & 6122$\pm$721 & 0.74$\pm$0.27\\
1430714491754808832 & 1430714496049436928 & 9.24 & 18.84 & 14791$\pm$5960 & 0.99$\pm$0.23\\
5198065190247097728 & 5198065228903571584 & 10.12 & 20.00 & 6931$\pm$892 & 0.75$\pm$0.27\\
5520033535248373120 & 5520033569597820544 & 8.69 & 20.03 & 7093$\pm$1076 & 0.94$\pm$0.24\\
\hline\hline
\end{tabular}
\end{table*}

\clearpage
\bibliography{DWD_crystallization}{}
\bibliographystyle{aasjournal}

\label{lastpage}
\end{document}